# Head Impact Characterization and Cellular Response of a Live-neuron cell-integrated Biomechanical Full-body Surrogate Model

Raisa Akhtaruzzaman[1], Mohammad Ibrahim Hossain[1], Rahid Zaman, Ashfaq Adnan*,

Mechanical and Aerospace Engineering Department, College of Engineering, The University of Texas at Arlington, Arlington, TX

## ABSTRACT

In this study, we develop a novel integrated framework that links the impact response with cellular dynamics using a live-neuron cell-integrated biomechanical full-body surrogate model. The impact event is simulated by allowing the surrogate model to fall from controlled seated release angles of 30°, 60°, and 90°. Three vertically stacked cell-culture Petri dishes, each containing live SH-SY5Y neuroblastoma cells, were placed inside the head of a commercially available surrogate model. The dynamic response of the impact event was evaluated using acceleration measurements from six accelerometers, comprising three sensors mounted on the head surface and three embedded in series with the cell stacks, along with kinematic measurements of the fall and deformation of the head model. In parallel, an OpenSim-based modified musculoskeletal model was used to simulate the fall experiment. We found that variation in contact stiffness produced the largest change in the predicted head acceleration in the simulation. When the cellular response and the measured accelerations are compared, oxidative stress and cell viability showed trends consistent with the regional acceleration and angle of fall. At the 90º fall, where median peak linear accelerations ranged from 170-258g, and the maximum headform deformation was approximately 9.4 mm, oxidative stress increased to approximately twice that of the control sample. We also quantified the cellular drift of SH-SY5Y cells, which is focal in nature for the 90º impact condition. The corresponding fall scenarios were also simulated in OpenSim and a preliminary calibration relationship was developed to compare the kinematic responses of the physical surrogate and musculoskeletal model. Finally, the framework provides a basis for relating experimental surrogate measurements to human head-neck response during impact.

* Corresponding author: aadnan@uta.edu, ashfaqadnan@gmail.com
[1] contributed equally.

# 1 INTRODUCTION:

Traumatic brain injury (TBI) is a major cause of death and disability worldwide [1]. It has a socioeconomic impact of more than $60 billion in the United States only, including hospitalizations and productivity losses [2,3]. TBI occurs when the head experiences external mechanical forces such as direct impact against a rigid surface (falls, vehicle crashes, sports collisions), being struck by or against an object (assaults, industrial accidents), penetrating injury from projectiles or sharp objects (e.g., bullets, shrapnel), rapid acceleration-deceleration of the head with or without impact (motor vehicle whiplash, shaken baby syndrome) or exposure to blast waves in military or civilian explosions. However, falls account for about 50-70% of work-related TBIs [2,4] and about 60% of TBIs in older adults [5]. A biomechanical study of workplace falls reported that 41% of backward falls and 19% of trips or forward falls produced head impacts outside the region of helmet coverage [2]. An important limitation in investigating falls is the inability to directly measure the impact forces experienced by an individual during a backward fall. Quantifying these forces is important for identifying potential injury severity. However, injury severity depends not only on the magnitude of the impact, but also on the direction of motion, impact location [6], contact surface, and response of the head and neck.

Fall studies have employed a variety of techniques to estimate impact forces and velocities, but these quantities cannot be directly measured during real-life injury events. In laboratory settings, low-impact falls may be measured using force plates, but for safety, impact energies or forces are maintained below injury-level conditions. Video-based fall analysis can provide kinematic information; however, additional methods are needed to characterize the forces associated with injury-level impacts. Because such forces cannot be measured directly in living subjects, controlled physical head models have been used to study how external impact loading is transmitted through the head. For example, force transmission through an anatomical head model has been quantified using an array of accelerometers [7]. Anthropomorphic test devices (ATDs), such as Hybrid III and THOR, are widely used in automotive and impact studies, but they provide limited insight into internal brain injury. In contrast, post-mortem human subjects (PMHS) provide high-fidelity anatomy but have limited availability and repeatability.

Computational multibody models have also been used to reconstruct head motion and injury during falls. Doorly and Gilchrist used a MADYMO multibody dynamics model to reconstruct ten real-world accidental falls that resulted in traumatic brain injury and reported reconstructed peak linear head accelerations ranging from 236.5 to 366.5g [8]. Non-contact head loading caused by sudden body motion has also been studied using finite element models to predict brain deformation, intracranial pressure, and brain stress [9]. The head may also experience rapid motion because of sudden vehicle shocks, and subject-specific biodynamic models have been used to study how these loads are transmitted from the seat through the upper body to the head [10].

Another aspect of impact testing is selecting the parameters that adequately describe the impact loading. Acceleration alone may not fully characterize the dynamics of an impact event. Previous studies have therefore considered impact speed, impact angle, and impact location, including whether the impact occurs within the helmet coverage area, and have used impact velocity as an indicator of impact severity. Based on a multibody simulation, it was found that median head impact speeds reached $8.1ms^{-1}$ for forward falls and $9.5ms^{-1}$ for backward falls from a height of 4 $m$, while trips without arm bracing produced a median impact speed of approximately $4.3ms^{-1}$ [2]. These findings show that impact velocity and fall configuration can vary substantially between different fall scenarios and should be considered alongside acceleration when characterizing head impact loading.

Musculoskeletal models can be used to study how impact location, impact severity, and neck muscle behavior affect head and neck motion. A head-neck musculoskeletal model showed that impact location and severity can significantly influence head-neck biomechanical responses, including neck force, neck moment, and injury criteria [6]. Along with that, the coefficient of restitution (COR) can be used to characterize rebound behavior and energy loss during high velocity head-surface impacts, including secondary rebounds caused by the head-neck motion. Furthermore, contact duration and peak deceleration can be investigated alongside COR to provide a more complete description of the impact event. Viano and Parenteau used COR to estimate peak head velocity in cadaveric head-impact tests and combined these data to examine relationships between head velocity, impact force, and serious neck compression injury [11]. Although impact kinematics and COR describe the motion before and after contact, they do not fully explain how the incoming motion is converted into contact force and deformation. The response depends on the contact geometry and the mechanical properties of the interacting bodies. Therefore, to define the impact characteristics of head-surface impacts, one key parameter is contact stiffness. Hertz contact theory provides a fundamental elastic relationship between normal contact force and indentation for two smooth curved spherical bodies in contact [12]. Contact stiffness formulation includes elastic behavior and viscoelastic behavior. The deformation behavior is sensitive to the mechanical stiffness of both contact materials (core and shell or shell and shell), and the relative stiffness plays a crucial role in determining how the system responds to external forces. However, the Hertz model does not account for energy loss during impact. The Hunt-Crossley model extends the Hertzian contact formulation by adding a nonlinear damping term to represent energy dissipation and differences between loading and unloading behavior [13]. The deformation of a spherical core-shell system depends on the relative stiffness, compressibility, and dimensions of the core and shell [14]. Although that study used a different type of loading than a head impact, it showed that deformation of the outer shell is affected by deformation of the inner core. Another study examined an elastic spherical shell pressed against a flat rigid surface and discussed how the shell flattens, deforms under load, and may buckle [15]. These studies provide a mechanical basis for examining how the foam core, outer skin, and rigid impactor jointly influence the contact response of the surrogate head. Because the surrogate may experience layered, viscoelastic, and relatively large deformation, Hertz theory is treated as an initial elastic model rather than a complete description of its contact behavior. The resulting contact response is also important for the internal mechanical environment of the headform. Contact stiffness, deformation, and energy dissipation at the head-surface interface can influence the local acceleration and loading experienced at locations inside the head. Therefore, understanding how the external impact is transferred from the contact surface to the internal region provides a direct link between head contact mechanics and cellular response.

Understanding how this internally transmitted mechanical loading affects brain cells is another challenge in studying fall-related TBI, as TBI is a complex, multiscale pathophysiological process. Linear axial impact tests on the SH-SY5Y human neuroblastoma cell line have been assessed in relation to linear acceleration [16]. Koumlis et al. used an impact model to apply linear impact using a spring-loaded test setup to apply controlled strain on live neuron cells and proposed an injury threshold [17]. Computational and molecular studies have also been used to investigate cellular and subcellular responses associated with mechanical brain injury [18], [19], [20], [21]. However, most previous studies either apply mechanical loading directly to isolated cell cultures or investigate cellular injury through computational models. They do not reproduce the sequence in which an external head impact is transmitted through a surrogate head and produces a local mechanical environment experienced by cells located inside the head. One potential approach to address this limitation is to incorporate living cells within an instrumented surrogate model and evaluate their biological response together with the local mechanical loading. Cellular injury can be assessed through several biological markers. In SH-SY5Y cells, exposure to mycotoxins can induce cytotoxicity in neuronal cells, accompanied by increased reactive oxygen species (ROS) and oxidative stress [22]. Theoretical and experimental studies have shown that mechanical loading can alter cell morphology and cytoskeletal organization through mechanosensitive cell-substrate interactions [23].

Despite extensive work on TBI biomechanics and pathophysiology, limited work has directly linked controlled, fall-relevant impact parameters (contact force profiles, impact and rebound kinematics, and resultant linear accelerations) with quantitative measures of neuronal cell response. Our study addresses this gap by experimentally characterizing brain cell responses under well-defined impact conditions, allowing the relationship between mechanical loading and cellular response to be evaluated. The surrogate head used in the study contains a compliant foam core covered by an outer skin and therefore cannot be treated as a homogeneous elastic sphere. In this study, cellular response following impact is evaluated using assays of ROS, cell viability, and morphological changes. The objectives of this study is : (1) experimentally characterize the impact dynamics of a seated physical surrogate model equipped with sensors and neuron cells under controlled release angles, (2) quantify the material and contact behavior of the surrogate head and neck, including stiffness, damping, and energy loss and calibrate the contact law parameters, (3) simulate equivalent seated release scenarios using an OpenSim musculoskeletal upper body model to estimate the human head-neck kinematics under these scenarios, (4) to develop a novel prototype framework that map the surrogate measurements to musculoskeletal response to neural damages, focusing on traumatic brain injuries. By connecting impact kinematics, contact mechanics, surrogate response, musculoskeletal simulation, and cellular outcomes, this novel framework aims to identify the mechanical parameters that control the surrogate response and improve the translation of surrogate measurements to human head and neck motion.

## 2 METHODS:

In this study, we introduced a novel multiscale framework to study traumatic brain injuries during controlled fall events. The framework consists of three modules: (1) a physical surrogate impact module, (2) an intracranial cellular response module, and (3) a musculoskeletal simulation module. The physical module measured the external impact response of the surrogate, the musculoskeletal module simulated the corresponding human head and neck motion, and the cellular module evaluated early biological responses to the impact.

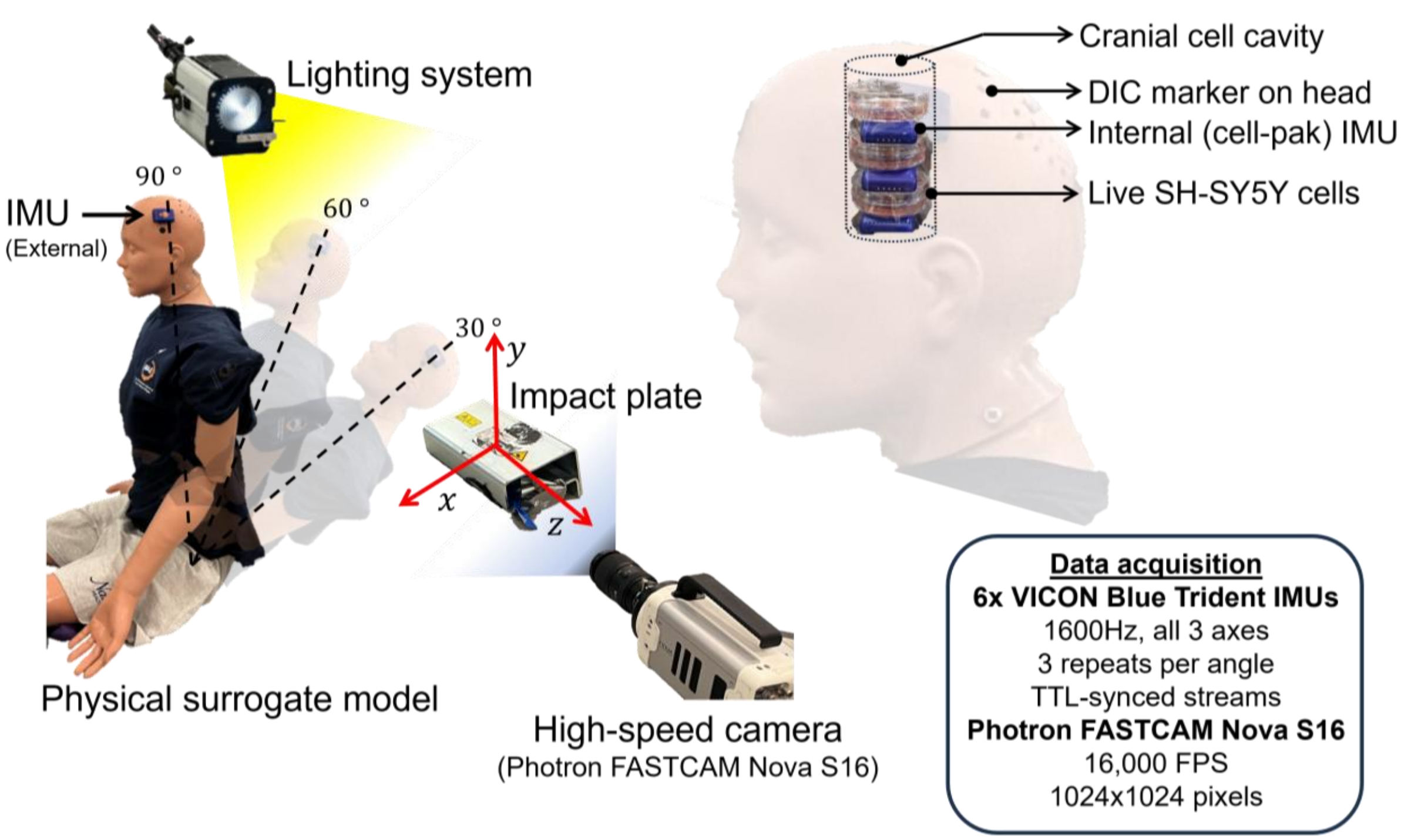

*Figure Schematic of the controlled surrogate fall impact setup showing the tested fall angles, internal and external IMU locations, intracranial cell pack positions, steel impactor plate, high-speed camera measurements, and data acquisition system.*

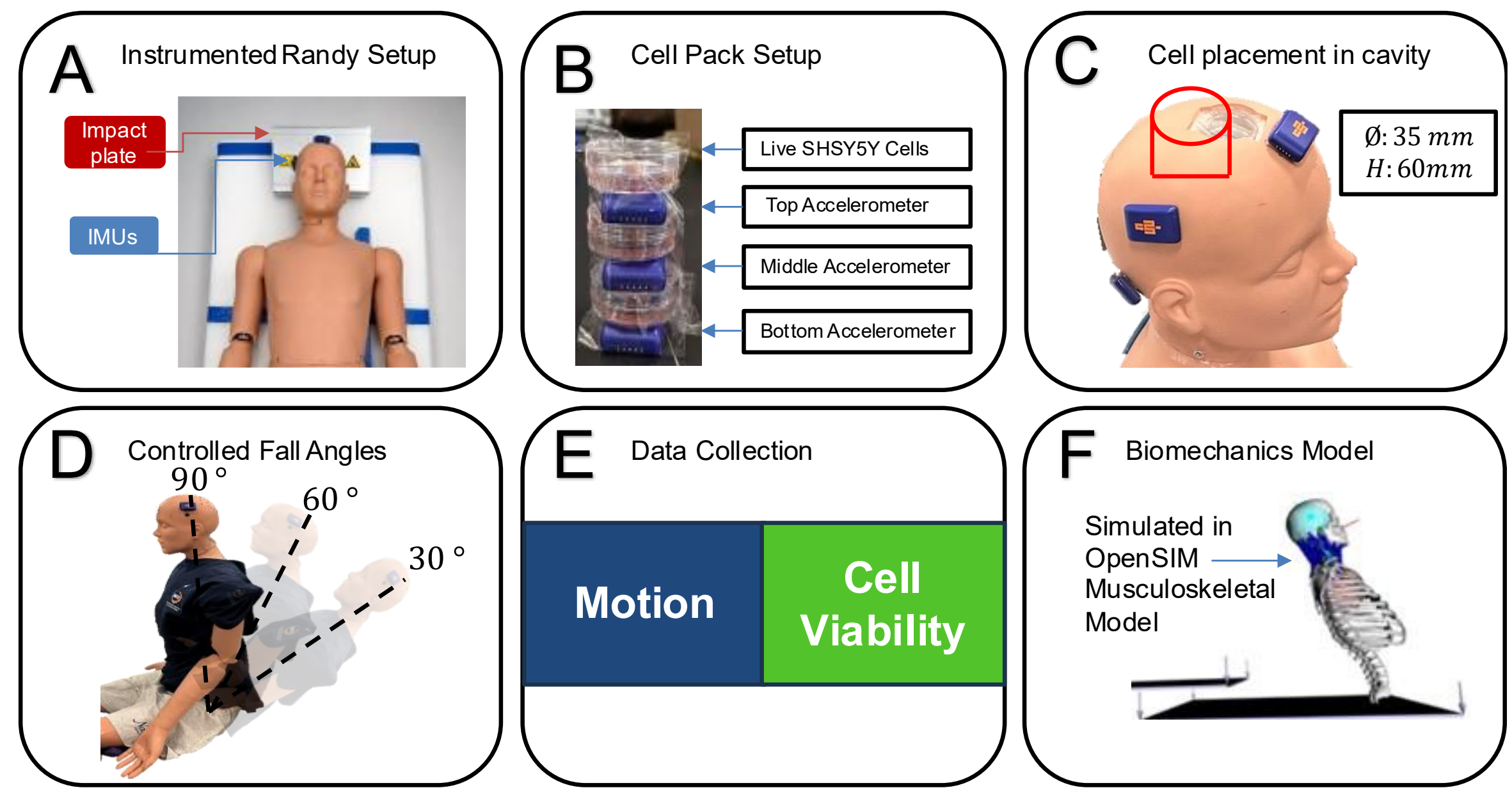

*Figure 1 Overview of the multiscale surrogate fall impact framework, including the instrumented surrogate setup, cell pack arrangement, intracranial placement and OpenSim musculoskeletal modeling.*

## 2.1 PHYSICAL SURROGATE IMPACT MODULE

### 2.1.1 SURROGATE AND FALL SETUP

A 165-lb physical human surrogate by Global Technologies, Davie, FL, USA was used to investigate head-impact kinematics, head deformation, rebound behavior, and head-neck motion during controlled seated falls. The experimental measurements were later compared with the musculoskeletal model predictions and cellular responses. The experimental framework was designed to relate measured head-impact kinematics to cellular responses and musculoskeletal model predictions. The objective is to find systematic patterns (e.g., higher peaks at specific fall angles) that can later be correlated with cellular viability outcomes. The complete experimental setup is shown in Figure  and Figure 1. As shown in Figure 1(D), the surrogate was released from seated configurations corresponding to fall angles of 30°, 60°, and 90° and was allowed to fall freely onto a rigid metal plate. The fall angle was defined as the angle of the surrogate torso relative to the horizontal ground plane, and a 3D-printed angle jig was used to establish the starting positions. For the 90° fall, no external push was required to initiate the motion. This behavior was attributed to the mechanical response of the surrogate hip hinge, which provided an initial moment that assisted the backward rotation of the upper body. A force of approximately 260 N was measured at the hip mechanism using a force meter. Using a hip-to-head distance of 0.838 m as the moment arm, the corresponding moment was estimated to be approximately 218 N·m.

### 2.1.2 HEADFORM MODIFICATION AND CELL-PACK PLACEMENT

The first modification to the surrogate was the addition of a multilayer cell pack containing SH-SY5Y cell cultures within the headform. A 35 mm diameter opening and a rectangular sensor slot were created in the superior region of the surrogate headform to accommodate the internal cell pack. The modified cavity had a depth of approximately 60 mm. The cell pack consisted of three Petri-dishes containing SH-SY5Y neuroblastoma cell cultures and three blue trident inertial measurement units (IMU) positioned at the top, middle and bottom of the cell pack. The arrangement and orientation of the Petri dishes and internal sensors are shown in Figure  and Figure 1. This configuration allowed the cellular response to be evaluated following the fall tests through changes in cell morphology, ROS production, and cell viability.

### 2.1.3 IMU INSTRUMENTATION AND DATA COLLECTION

A total of six Inertial Measurement Unit Sensors (IMU's) by Vicon Motion Systems Ltd., Oxford, UK were used to measure the head kinematics of the surrogate during each fall. The sensors recorded triaxial motion along the X, Y, and Z axes at a sampling frequency of 1600 Hz. These accelerometers have a range of ±200g in each axis in "high g" mode. Along with the three sensors that were placed inside the cell pack, the remaining three sensors were attached to the external surface of the headform at the forehead, parietal, and occipital regions, and all six sensors recorded motion data simultaneously. The external sensors were secured using double-sided adhesive tape to minimize relative motion between each sensor and the headform. An additional IMU sensor was used for velocity measurement as in "high g" mode these sensors cannot give velocity measurements. Figure  and Figure 1 show the locations and orientations of the sensors and Petri dishes within the cranial cavity. Before testing, the sensors were calibrated according to the manufacturer's recommended procedure, and each fall condition was repeated three times.

### 2.1.4 HEAD IMPACT AND CONTACT ANALYSIS

The head impact on the steel plate is analyzed using the classical Hertz contact theory [12]. The theory provides a classical analytical framework for describing the local deformation and contact force between two elastic bodies brought into contact. The Hertzian contact force F is related to the collective deformation of the contacting bodies, $\delta$ by,

$$F = \frac{4}{3} E^* \sqrt{R^*} \delta^{\frac{3}{2}} \tag{1}$$

Here, $E^*$ represents the effective Young's modulus, $\delta$ represents the deformation, and $R*$ represents the effective radius of the contacting bodies of radius $R_1$ and $R_2$ by the following relation:

$$\frac{1}{R^*} = \frac{1}{R_1} + \frac{1}{R_2} \tag{2}$$

The effective elastic modulus is defined as

$$\frac{1}{E^*} = \frac{1 - v_1^2}{E_1} + \frac{1 - v_2^2}{E_2} \tag{3}$$

Where, $v_1$ $and$ $v_2$ are Poisson's ratios of contact materials. For our experiment, one of the bodies is the flat steel plate that can be approximated as a body with large radius and stiffness. Then, $R^* \approx R_1$ and $E^* = \frac{E_1}{1-{v_1}^2}$. To apply the contact theory, the parameters $R_1$ and $E_1$ are needed.

A Photron FASTCAM Nova S16 high-speed camera (Photron, San Diego, CA, USA) was used to record the head motion and surface deformation during impact. High-contrast markers were placed at selected locations on the headform and tracked throughout the impact event. The camera recorded the motion at 16,000fps with a resolution of 1024x1024 pixels. The displacement of each marker was calculated relative to a fixed reference point, and differences in marker displacement were used to estimate deformation of the headform. The differential displacement between a marker located near the impact region and a second marker located away from the impact region was used to quantify the local deformation. The maximum differential displacement was taken as the peak deformation, while the slope of the compression portion of the displacement–time response was used to estimate the deformation velocity.

The volume of the headform undergoing deformation and the contact area during impact were also estimated from the high-speed imaging measurements. The measured deformation was subsequently used to estimate the deformation energy, which was compared with the change in translational kinetic energy across the primary impact and rebound event. To further characterize the mechanical response of the surrogate headform, compression tests were performed using a universal testing machine at loading rates of 1, 10, 100, and 1000 mm/min. The force–displacement response from these tests was used to determine the effective material properties of the headform. A characteristic length of 160 mm and a characteristic area obtained from the paint-based contact measurement were used in the material-property calculation. Using these dimensions together with the slope of the force-displacement response, the effective modulus was calculated as follows:

$$E_{SURROGATE\ HEAD} = E_1 = \frac{\partial F/A}{\partial l/l} \tag{4}$$

Here, $F$ is the impact force, $A$ is the contact area, and $l$ is the characteristic length. Because the force-displacement response varied with loading rate, a rate-dependent power law relationship was used to describe the effective modulus as a function of compression rate:

$$E_1 = C\dot{\delta}^m \tag{5}$$

where $E_1$ is the effective modulus of the surrogate headform, $C$ is the fitted material coefficient, $\dot{\delta}$ is the compression rate, and $m$ is the rate-dependent exponent.

To calculate Poisson's ratio, a section of the surrogate headform was cut into a square sample and compressed at a loading rate of 1000 mm/min, which was the highest loading rate available from the universal testing machine. During the compression test, a Photron NOVA S-series high-speed camera was used to capture the lateral expansion of the material under axial compression. The axial deformation was obtained from the force–displacement test data, while the lateral deformation was obtained from the image-based measurements. Using these two-deformation data, Poisson's ratio was calculated. The experimentally determined modulus and Poisson's ratio were then used to calculate the effective contact modulus.

The radius of curvature of the head was estimated using measuring tapes. The experimentally determined contact stiffness was subsequently used in the Hunt-Crossley contact formulation of the OpenSim model. Relative head-neck motion (whiplash effect) was also evaluated from the high-speed camera measurements. These experimental results were later compared with the head–neck response predicted by the OpenSim musculoskeletal model.

## 2.2 INTRACRANIAL CELLULAR RESPONSE MODULE

Human neuroblastoma SH-SY5Y cells were obtained from ATCC, CRL-2266. Cells were maintained at 37 °C in a 5% CO2 humidified incubator. Prior to the experiment, cells were cultured in DMEM (D5796, Sigma Aldrich) culture medium supplemented with 10% FBS (10,437,028, ThermoFisher), 100 g/mL penicillin streptomycin (15140122, ThermoFisher) and 2 mM L-glutamine (A2916801, ThermoFisher). For further experiment, upon reaching 80% confluence, the cells were digested with 0.25% trypsin for 3 min until cells are detached. The cells were subsequently resuspended with a DMEM culture medium containing 10% FBS. All steps were taken in a sterile laminar cabinet. Following centrifuging at 1000g for 2.5 min and cells were resuspended with fresh culture medium. Just after impact the cells were taken into the BSL-2 cabinet to add LIVE/DEAD assay kit for fluorescence staining. The cells were taken into the incubator and kept for 20 minutes. Then, the cells were taken inside the microscope incubator and imaged. The cells were images approximately 25 minutes for after impact event capture.

Cell oxidative stress was observed using a BZ800X fluorescence microscope each day (Keyence, USA). The fluorescence excitation maximum is shifted to a specific wavelength 560/40nm for red fluorescence, and proper optical filters (TxRed) were used with a magnification of 20 x. The images are post-processed using both BX analyzer and ImageJ for image clarity and morphological change detection, etc. Cell counting was performed using a microplate reader (Invitrogen, ThermoFisher Scientific, USA) on days 2, 4, and 6 of the culturing period. SH-SY5Y cells were plated in a 35mm petri dish (Corning Costar) with 300,000 cells per ml in media. To perform the analysis, cells were stained, and a drop test was done before that. After the experiment, images were captured of the cells. The cells emit red fluorescence according to their state of viability. To label Reactive Oxygen Species, we used ROS assay kit (L34977, Invitrogen). Additionally, the viability of the cells was assessed using an MTS assay kit (G3580, Promega) after 1 hr, and cell viability was calculated based on the absorbance of the sample at a wavelength of 490nm. Briefly, SH-SY5Y cells were plated in a 96-well plate (655,098, Fisher) with 10,000 cells per well in 100 µl MTT (0.5 mg/ml) media. 1:5 ratio of media and MTS reagent was mixed. The plate was incubated at 37 °C for 4 h before the MTT absorbance was read at OD 490 nm by Molecular Devices microplate reader (Molecular Devices, LLC). The cell proliferation assay was performed in triplicate for each group in each set.

## 2.3 MUSCULOSKELETAL SIMULATION MODULE

We used an in-silico approach to investigate the response of our surrogate system and identify potential mismatches between musculoskeletal model response and surrogate response. We used OpenSim environment in this study. OpenSim is an open-source software for biomechanical modeling, simulation, and analysis. The experimental fall conditions were reproduced in OpenSim using a modified upper extremity musculoskeletal model [24] to represent the human body with segment inertias scaled to physical surrogate anthropometry. The model was simulated for the three experimental fall conditions (30°, 60°, and 90°), with initial conditions obtained from the experimental measurements. We considered only the neck muscles in the musculoskeletal model, as the head response on the floor is mainly dominated by neck muscle response and torso muscles are not activated, as it is mostly free fall. The pelvis was connected to the ground via a 6-DOF joint. We allowed only the rotational 3 DOFs at the pelvis-ground joint so that the torso and head could undergo sagittal-plane backward rotation under gravity, analogous to the physical surrogate model from an initial angle and velocity. We run the forward dynamics simulations for all three falls (90º, 60º, and 30º) with input boundary motion conditions from the experiment to mimic the motion profile. We used the initial velocity of the surrogate that occurs due to the hinge pressure in addition to the free fall, and the Hertz contact stiffness is taken from the experiment.

The dynamics of the human musculoskeletal system can be formulated with the Euler-Lagrange equations [6] as per Eq. (6):

$$\mathbf{M(q)\ddot{q}} = \mathbf{C(q,\dot{q})} + \mathbf{G(q)} + \mathbf{R(q)F^{T}(u)} + \mathbf{E(q,\dot{q})} \tag{6}$$

where, $\mathbf{q}$, $\dot{\mathbf{q}}$, and $\ddot{\mathbf{q}}$ are the positions, velocities, and accelerations of the joints, $\mathbf{M(q)}$ is the mass distribution matrix which contains masses and inertial properties of the body segments, $\mathbf{C(q,\dot{q})}$ is the Coriolis and centrifugal force vector which arises when Newton's laws of motion are applied in reference frames that are fixed to rotating bodies, $\mathbf{G(q)}$ is the gravitational force vector, $\mathbf{R(q)}$ is muscle moment arm matrix, $\mathbf{F^{T}(u)}$ is the tendon force vectors, which is a function of muscle excitations ($\mathbf{u}$), $\mathbf{E(q,\dot{q})}$ is the external forces vector that represents the interactions between the body and environment.

Head-plate and torso-floor interactions were modeled using two Hunt-Crossley contacts element between a spherical contact patch rigidly attached to the skull (and torso), and a planar half-spaced attached to the steel plate. The normal contact force $f_n$ can be obtained from Hunt-Crossley law:

$$f_n = kx^n(1 + c\dot{x}) \tag{7}$$

where, $k$ is the stiffness constant incorporating material properties and geometry, $c$ is an effective dissipation coefficient, $x$ is the penetration depth, $\dot{x}$ is the penetration rate (positive during penetration and negative during rebound). $n$ depends on the surface geometry, for sphere, $n = 3/2$. The first term ($kx^n$) has been quantified in the experiment from Hertz Contact Theory and has been used in the OpenSim model which is equal to $2 \times 10^6$ $N.m^{-1.5}$ [25].

During the loading (head moving towards the floor, $\dot{x} > 0$), $(1 + c\dot{x}) > 0$. So, force is higher than pure Hertz. As a result, more work is done, and some is converted to heat or viscoelastic loss. During unloading, $(1 + c\dot{x}) < 1$. So, force is lower and less elastic energy is recovered. So, increasing the dissipation coefficient will lower the coefficient of restitution and will have less rebound.

The tangential friction force magnitude is:

$$f_t = f_n\left[\min\left(\frac{v_s}{v_t}, 1\right)\left(u_d + \frac{2(u_s - u_d)}{1 + \left(\frac{v_s}{v_t}\right)^2}\right) + u_v v_s\right] \tag{8}$$

Where $f_n$ is the normal contact force (from Hunt-Crossley), $v_s$ is the slip speed at the contact, $v_t$ is the transition slip speed between static and dynamic friction, $u_s$ is the static friction coefficient, $u_d$ is the dynamics friction coefficient, and $u_v$ is the viscous friction coefficient. Details about the normal and tangential contact forces can be found in [26,27] .The contact details used in this simulation are provided in Table 1. The simulation reproduced three experimental conditions in which the seated surrogate model was released from different initial torso angles relative to the horizontal ground plane: 90°, 60°, and 30° as mentioned above. In the model, these were implemented as different initial generalized coordinates of the pelvis-to-ground joint, with small initial angular velocities chosen to match the measured initial condition and spring-derived response at release of the surrogate model. For each condition, the system was then allowed to evolve freely under gravity with an initial velocity of 20 deg/s based on the experimentally observed initial motion of the surrogate. Head kinematics at the impact location were recorded for comparison with accelerometer data. The experimentally determined Hertzian contact stiffness was used directly in the Hunt-Crossley contact model. Literature values from computational fall studies were used to guide the selection of floor-contact properties [28]. The complete set of contact parameters is provided in Table 1.

*Table 1 Head-floor and torso-floor contact details*

| | Head -floor contact | Torso-floor contact |
|---|---|---|
| $k$ | $2x10^6$ | $8.6x10^6$ |
| $c$ | 0.5 | 0.5 |
| $v_s$ | 0.9 | 0.9 |
| $u_d$ | 0.9 | 0.9 |
| $u_v$ | 0.6 | 0.6 |
| $v_t$ | 0.1 | 0.1 |

# 3 RESULTS AND DISCUSSION:

The results are presented in four parts to describe how the impact response develops from whole-body motion to cellular-level effects. First, the kinematic response of the surrogate is evaluated using the internal and external IMU measurements, with emphasis on the effects of fall angle and sensor location on linear acceleration. Second, the cellular response is assessed using the multilayer SH-SY5Y cell pack by comparing cell morphology, reactive oxygen species (ROS), and cell viability before and after impact. Third, high-speed imaging and marker tracking are used to quantify headform deformation and evaluate the relationship between impact energy and deformation. Finally, the biomechanical analysis compares the experimental surrogate response with the OpenSim head-neck model and examines the contact parameters used in the Hertz and Hunt-Crossley formulations. Together, these four analyses connect the external impact conditions, surrogate-head response, musculoskeletal motion, and early cellular changes within a single multiscale framework.

## 3.1 KINEMATIC RESPONSE

At the 60° and 90° falls, the cell-pack sensors generally recorded higher peak resultant accelerations (with median peak acceleration of 241.69 g at 60° and 257.8 g at 90°), than the sensors placed on the external surface of the headform. In contrast, the regional responses were in close proximity to each other at 30°, with peak median accelerations ranging approximately from 51.58-99.06 g across the six sensor locations (Figure 2). This may be related to the location of the cell pack near the primary impact path and the local motion of the internal assembly. The differences in regional acceleration also increased with increasing fall angle, indicating a less uniform head response at higher fall angles.

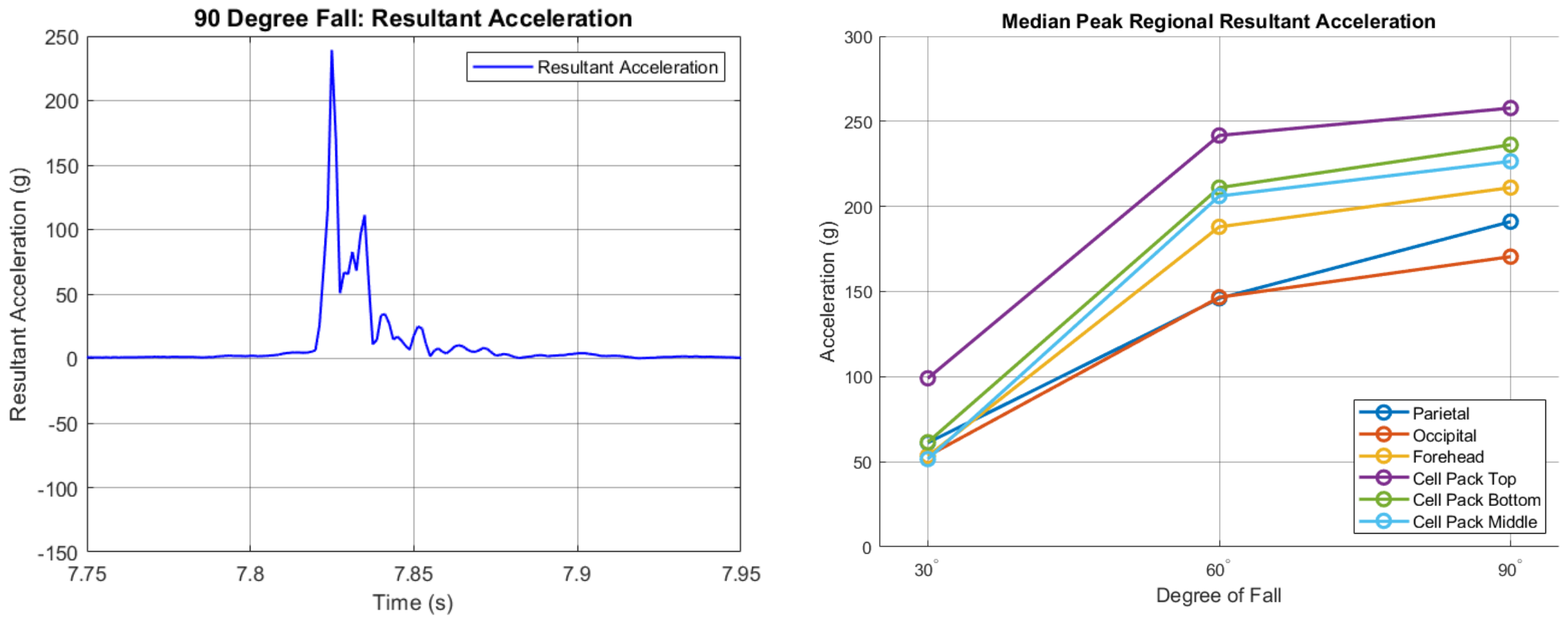

*Figure 2 (a) Resultant acceleration response of the bottom cell-pack IMU during the 90° surrogate fall. (b) Median peak resultant linear acceleration at the external head and cell-pack sensor locations for the 30°, 60°, and 90° fall conditions.*

The single-sensor acceleration profile in Figure 2(a) also shows that the impact response is not limited to a single acceleration peak. Multiple impact pulses are observed, with the peak magnitude generally decreasing after the primary impact. These later pulses likely correspond to secondary head-plate contacts after the initial rebound, as continued relative motion of the head and neck brings the head back toward the plate. For the regional comparison shown in Figure 2(b), the first major impact peak from each sensor acceleration profile was used, as the subsequent impact peaks were considerably smaller. Additional single-sensor acceleration profiles are provided in the Supplementary Information. Overall, the acceleration response varies across the headform because of differences in sensor location, contact conditions, headform geometry and material properties, and the coupled motion of the head and neck. The regional peak accelerations ranged approximately from 51.58-99.06 g for the 30° fall, 146.04-241.69 g for the 60° fall, and 170.49-257.8 g for the 90° fall.

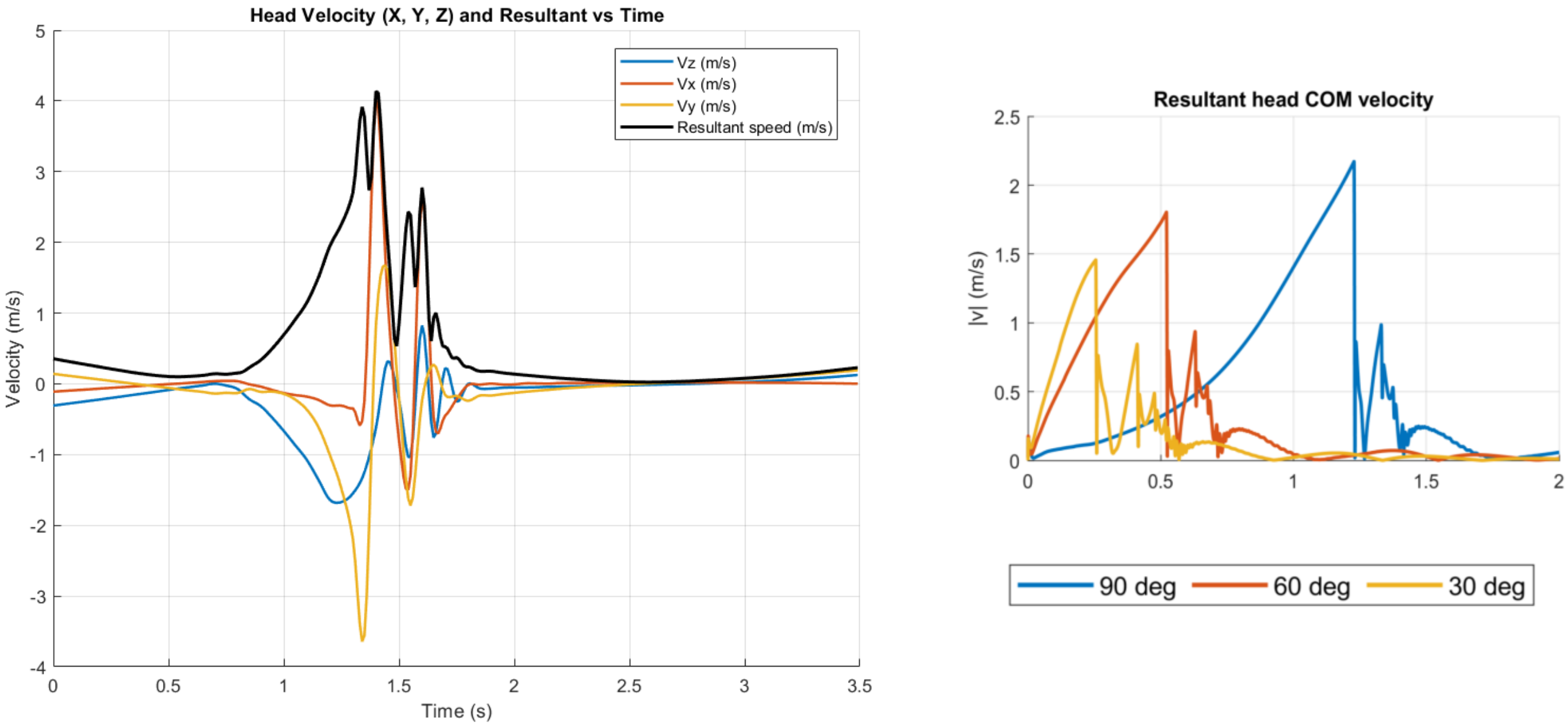

*Figure 3 a) Experimental head velocity components and resultant velocity during the 90° fall, showing the primary impact and rebound response. b) Resultant head center-of-mass velocity from the OpenSim simulations for the 30°, 60°, and 90° fall conditions.*

Figure 3 (a) shows the experimental head velocity response during the 90° fall. The velocity increases as the head approaches the plate, followed by a rapid change in magnitude and direction at the primary impact. The reversal of the impact direction velocity after contact represents the rebound of the surrogate head, while the subsequent smaller oscillations are consistent with the repeated impact behavior observed in the acceleration response. For the 90° fall, the peak velocity immediately before impact was approximately 3.92 $ms^{-1}$ in the impact direction, while the first rebound velocity was approximately 2.51 $ms^{-1}$. Figure 3(b) shows the corresponding resultant head center-of-mass velocity predicted by OpenSim for the 30°, 60°, and 90° fall conditions. The simulated velocity increased with increasing fall angle, consistent with the greater impact severity observed experimentally at the higher fall angles. The head velocities immediately before the primary impact and immediately after the first rebound were subsequently used to estimate the change in translational kinetic energy across the impact event. Here, the kinetic-energy difference refers specifically to the change between the instant immediately before primary head impact and the instant immediately after the first rebound, rather than the beginning and end of the complete fall event.

The translational kinetic energy immediately before the primary impact was defined as $KE_1$, while the kinetic energy immediately after the first rebound was defined as $KE_2$. The difference, $\Delta KE = KE_1 - KE_2$, represents the reduction in translational kinetic energy across the primary impact and rebound event. The corresponding $\Delta t$ represents the elapsed time between the instants at which $KE_1$ and $KE_2$ were evaluated and should therefore be interpreted as the impact rebound interval rather than the full duration of the fall. The reduction in translational kinetic energy increased with the severity of the fall condition. These kinetic-energy estimates are later compared with the experimentally estimated deformation energy in Section 3.3 to evaluate how the measured impact energy relates to headform deformation.

*Table 2 Pre impact kinetic energy ($KE_1$), post rebound kinetic energy ($KE_2$), kinetic energy difference (ΔKE), and impact rebound time interval (Δt) for the 30°, 60°, and 90° fall conditions.*

| 90° | 60° | 30° |
|---|---|---|
| $\Delta KE = KE1 - KE2 = 3.59J$<br>$\Delta t = 0.1s$ | $\Delta KE = KE1 - KE2 = 2.85J$<br>$\Delta t = 0.08s$ | $\Delta KE = KE1 - KE2 = 1.63J$<br>$\Delta t = 0.06s$ |

## 3.2 CELLULAR RESPONSE

Following the impact event, the response of the cells inside the surrogate head was evaluated to determine whether the mechanical loading experienced within the headform was associated with measurable changes in the SH-SY5Y cells. Cellular response was assessed through morphological observations, Live/Dead fluorescence imaging, reactive oxygen species (ROS), and cell viability measurements. The phase contrast images were first examined to identify visible changes in cell morphology following impact.

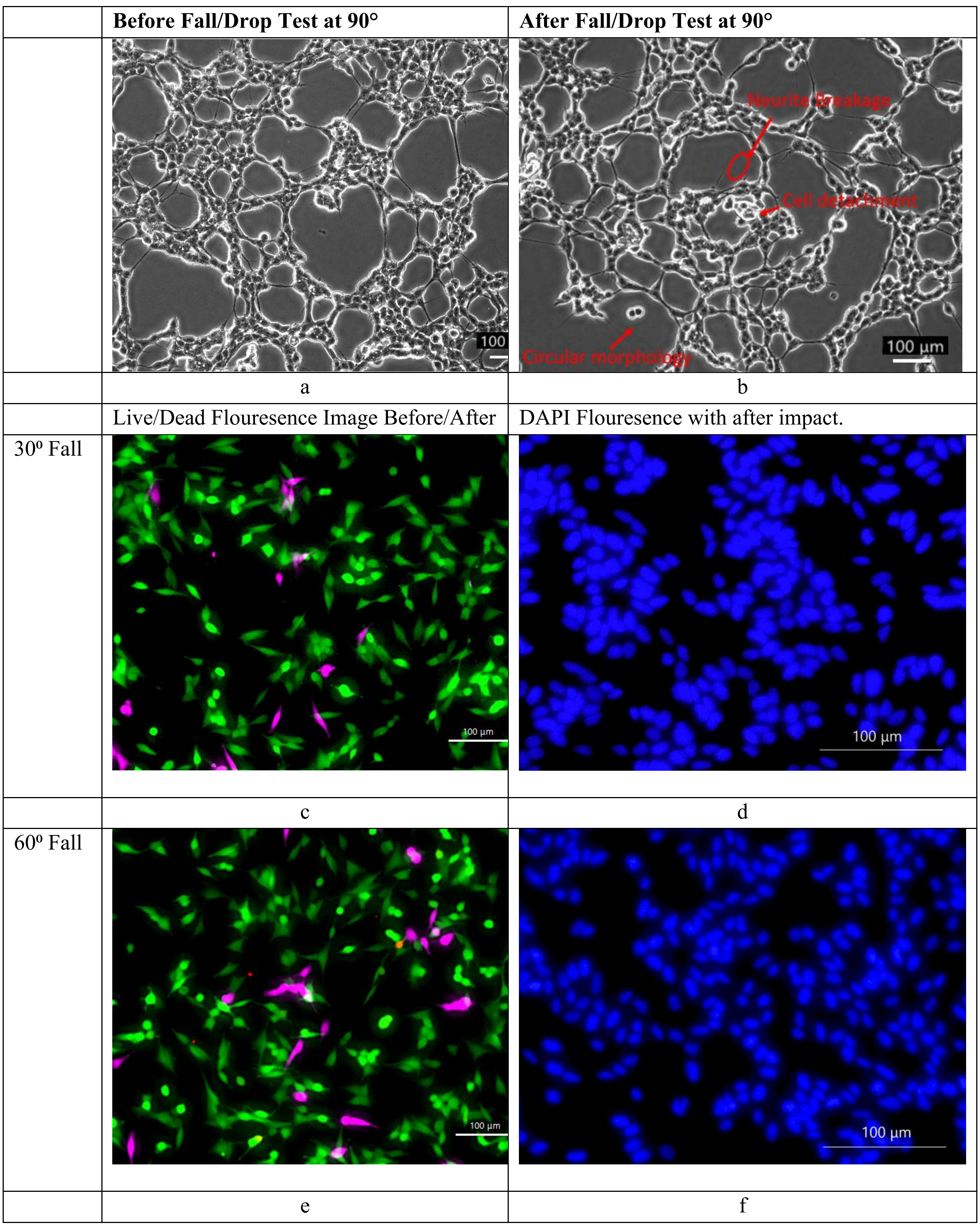

Before Fall/Drop Test at 90°
After Fall/Drop Test at 90°
100
Neurite Breakage
Cell detachment
Circular morphology
100 µm
a
b
Live/Dead Flouresence Image Before/After
DAPI Flouresence with after impact.
30º Fall
100 µm
100 µm
c
d
60º Fall
100 µm
100 µm
e
f

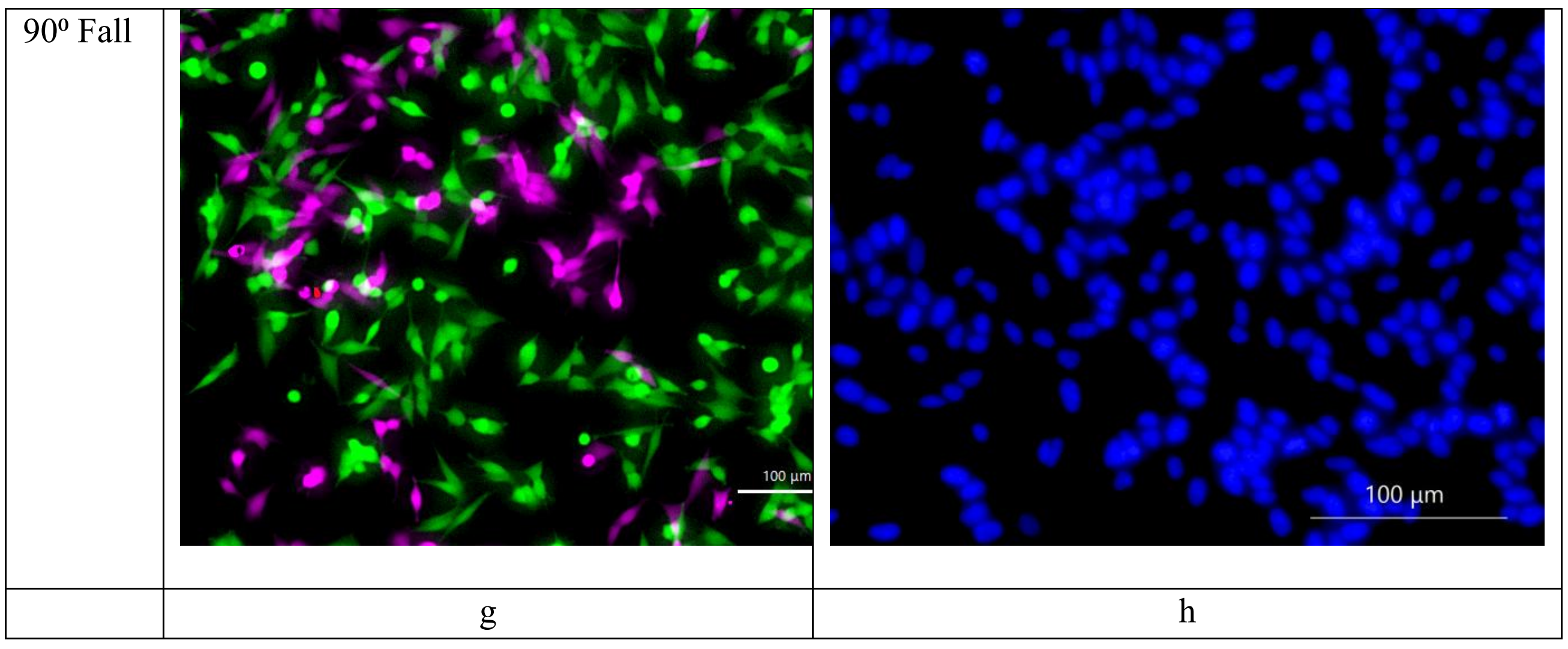

*Figure 4 Representative morphological and fluorescence images of SH-SY5Y cells following surrogate fall impacts. (a,b) Phase-contrast images before and immediately after the 90° fall, respectively. (c,e,g) Registered Live/Dead fluorescence images for the 30°, 60°, and 90° fall conditions, respectively, using Calcein AM and Ethidium Homodimer. In the registered images, purple regions indicate cells present before impact that were not observed at the same location after impact, whereas green regions indicate cells observed at similar locations before and after impact. (d,f,h) DAPI-stained fluorescence images obtained after the 30°, 60°, and 90° falls, respectively, showing cell nuclei. Scale bar = 100 µm.*

We observe distinct differences between the pre-impact and post-impact conditions for 90 deg fall, as shown in Figure 4(a,b). We observe some neurite breakage, indicating potential disruptions of communication between the cells. We also observe cell detachment following the impact induced by the high angle fall. Finally, the circular morphology of the cells further indicates loss of connection/communication between cells and cytoskeletal disruption due to induced mechanical stress; hence, these cells will eventually become metabolically inactive. Also, the 2D cells create ligand bonds with the tissue culture-treated Petri dish used in the experiment. We observe a portion of the cells drift away upon impact (purple-colored cells) (Figure 4c, e, g). Other neuron cells remain in the same position, with some SH-SY5Y cells dead, stained red by the LIVE/DEAD fluorescence dye used. The cell drift is more pronounced at the 90⁰ impact test, where the median peak resultant acceleration across the three cell-pack locations was approximately 236 g. Also, at 90⁰ impact, where the peak resultant linear acceleration reaches as high as 278g in the cell pack, the cell drift is clumped in the same location, inferring focal injury. Previous in-vitro studies of SH-SY5Y cells have shown that increasing mechanical strain can reduce cell viability, with the magnitude of the initial mechanical loading playing an important role in determining cellular injury severity [29]. In contrast, the drifted cells in 30º fall, with a peak resultant acceleration of approximately 99g, are more diffuse in nature, while the 60° condition showed an intermediate distribution. Also, DAPI fluorescence dye was used to stain the somas attached to the Petri dish and considered functional cells. A very slight decrease in the number of somas is observed in the image, though the difference is not significant. Cells that have undergone physiological pathways that affect cellular health, for example, the mechanical stress that is induced to the cells as a response to the cells exhibiting oxidative stress.

By staining the cells with an appropriate assay kit, we can visualize the oxidative stress in the form of red stains as shown in Figure 5. These responses, together with the metabolic activity of the cells, have been plotted and shown in Figure 6. From Figure 6(b, d, and f), we observe that the oxidative stress in the cells is quite low in the case of the 30° and 60° falls but quite prominent for the 90° fall. The corresponding median cell-pack peak accelerations were approximately 61 g, 211 g, and 236 g for the 30°, 60°, and 90° falls, respectively. For the 90° fall, the experimental resultant head velocity immediately before impact was approximately $3.92\ ms^{-1}$, for 60°, it was $3.31ms^{-1}$ and for 30° it was $2.84ms^{-1}$. On the other hand, the cellular viability of the cells is visible at all angles with minimal response at 30°. Increase in ROS means a cell's ability to neutralize or eliminate them through antioxidant mechanisms is falling. On the other hand, viability indicates how metabolically active cells are. Hence, viability results and Reactive oxygen species result show close harmony at different angles of fall tested. The acceleration of the neuron cells in all the three cell packs has been quantified using accelerometers attached to the petri-dishes as shown in Figure and Figure 1.

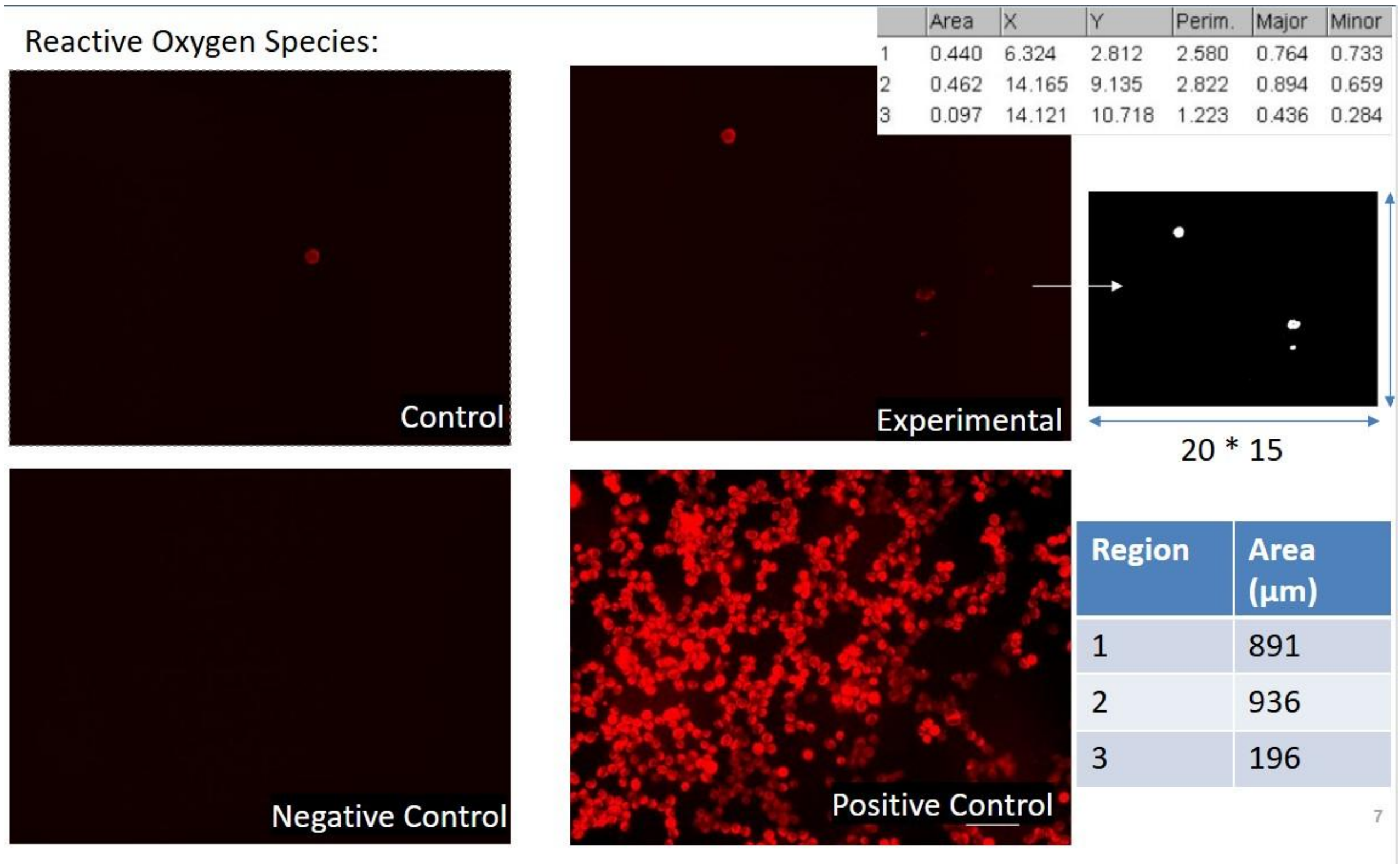

*Figure 5 Quantification of ROS from 20X fluorescence images of stained SH-SY5Y cells.*

From Figure 2, we generally see that the top cell pack experiences the highest acceleration compared to the other five locations. The second highest acceleration is the bottom cell pack acceleration, and the third highest acceleration is the bottom cell pack acceleration. If we compare only between the 3 cell packs, middle pack has the lowest acceleration. From Figure 6 we see that the ROS results of more accelerated cells at top location have highest ROS, and middle cell ROS is the lowest. This difference in result at different locations is more prominent at 90º fall as it experienced more acceleration. At lower accelerations e.g., 30º, the differences between the cellular responses are minimal. The viability of the cells decreases at a faster rate for top cell packs which experience a higher acceleration showing regional differences in cellular response at a given angle of fall.

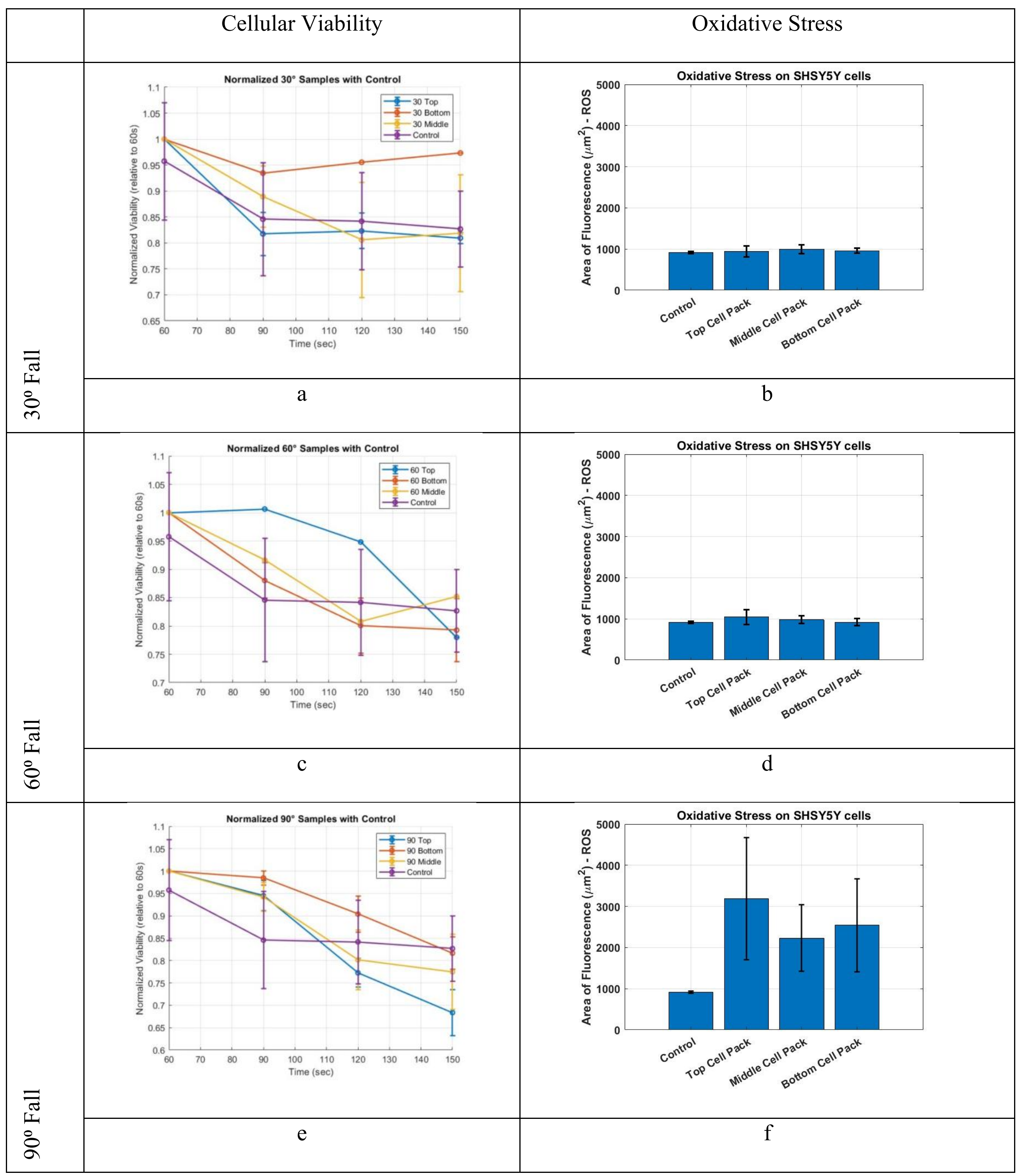

*Figure 6 Comparison of the Cellular Viability and Oxidative Stress and the Viability of the cells at different drop angles.*

The viability of the control sample initially drops as the cell sample is taken out of incubator and exposed to the environment without carbon dioxide and optimal temperature but stabilizes and remains same afterwards. In Figure 6e, from 60 to 150s, the normalized viability decreased from approximately 1.00 to 0.68 for the top cell pack, 0.82 for the bottom cell pack, and 0.77 for the middle cell pack, corresponding to decreases of approximately 32%, 18%, and 23%, respectively. By 150s, the average normalized viability across the three cell-pack locations was approximately 0.76 for the 90° condition, compared with approximately 0.81 for 60° and 0.87 for 30°, indicating a greater overall reduction in viability at the highest fall angle. Hence, we conclude that at approximately 200g and at 90 degree fall we start to see substantial cellular injury.

## 3.3 HEAD DEFORMATION ANALYSIS:

Using the high-speed imaging and differential-displacement procedure described in Section 2.1.4, the local deformation of the surrogate head was quantified during impact. Figure 7 shows the differential displacement response used to identify the peak deformation and compression velocity.

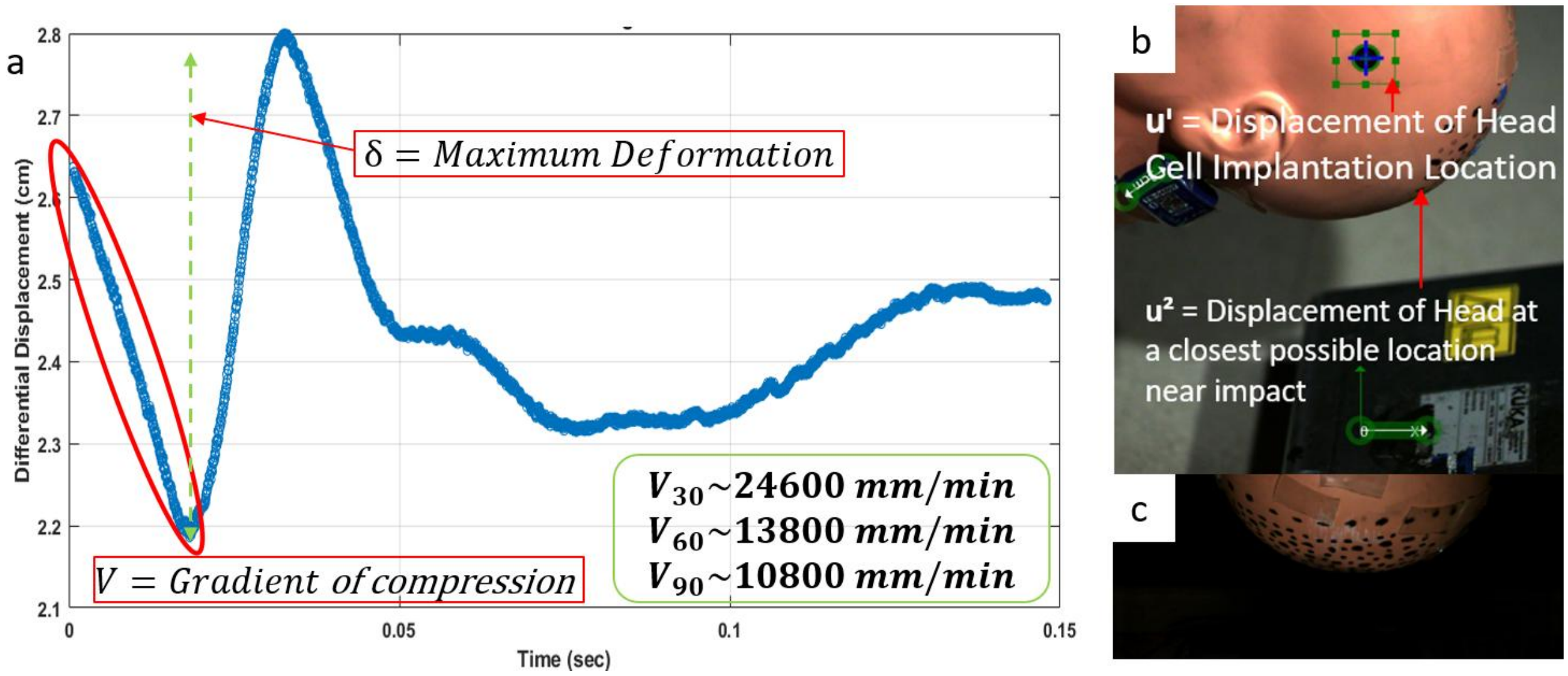

*Figure 7 Quantification of surrogate head deformation from high-speed imaging. (a) Differential displacement–time response used to determine the maximum deformation, δ , and compression velocity from the slope of the loading portion of the curve. The estimated compression velocities for the 30°, 60°, and 90° fall conditions are also shown. (b) Marker locations used to calculate the differential displacement between the cell implantation region and a location near the impact site. (c) High-speed image showing the headform tracking markers.*

The differential displacement response in Figure 7(a) shows a rapid compression of the headform during the initial impact followed by recovery and subsequent oscillations. The maximum deformation was obtained from the difference between the initial reference displacement and the minimum displacement reached during the compression phase. The corresponding compression velocities were approximately 24,600 mm/min, 13,800 mm/min, and 10,800 mm/min for the 30°, 60°, and 90° fall conditions, respectively.

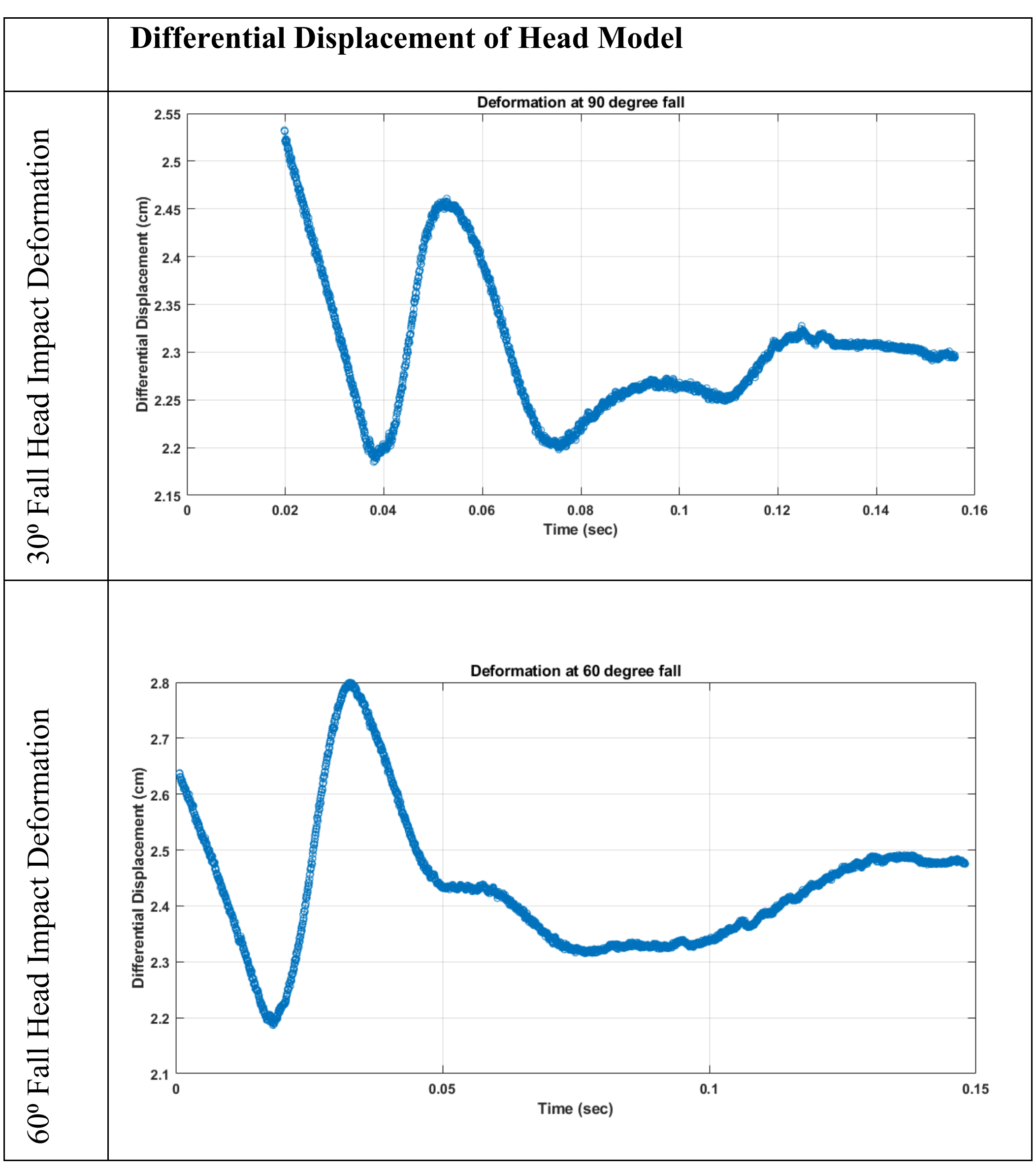

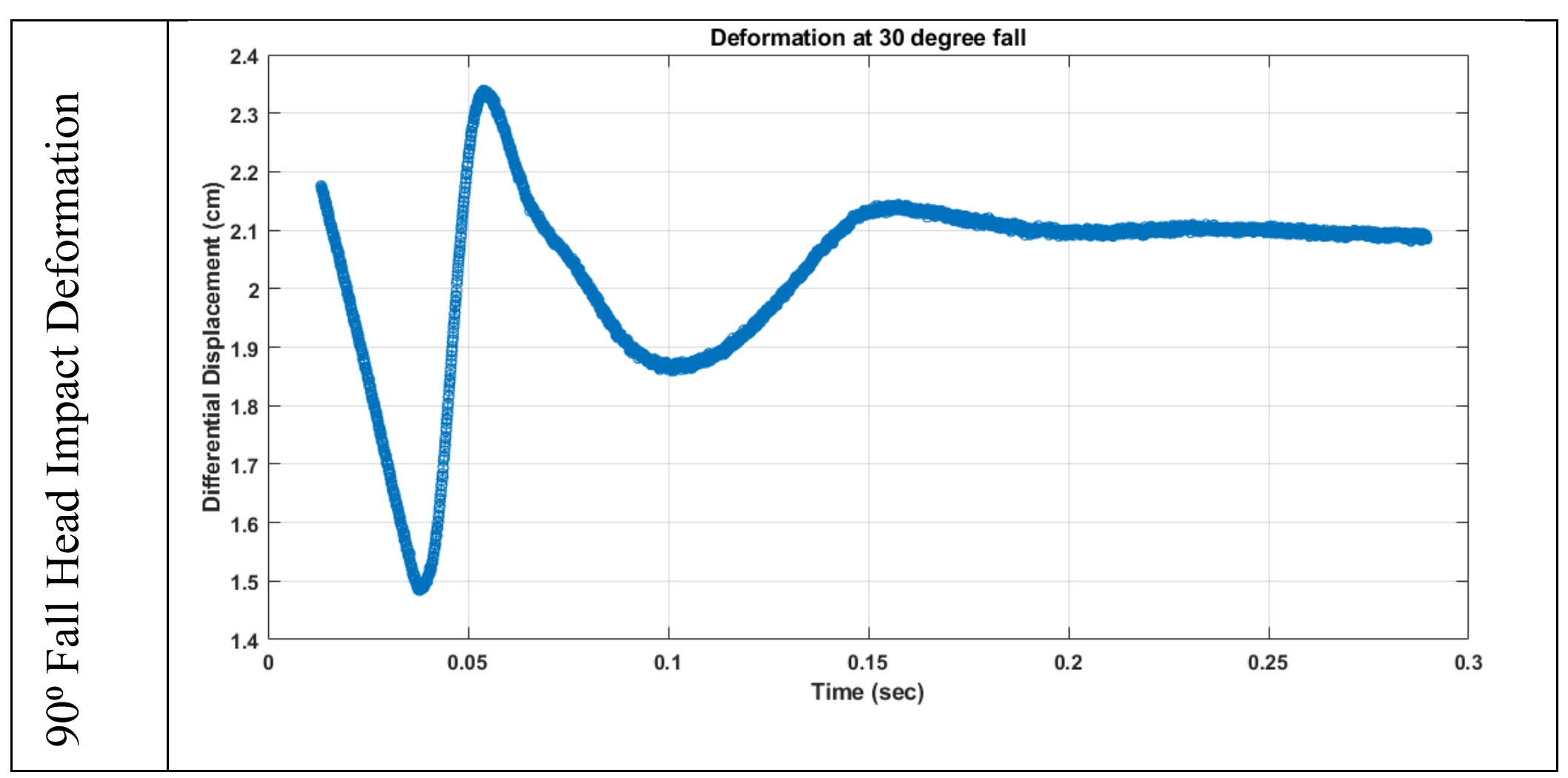

*Figure 8 Differential displacement histories of the surrogate headform during the 30°, 60°, and 90° fall conditions, showing the deformation response following head plate impact.*

Figure 8 compares the differential displacement response of the headform for the three fall conditions. The maximum measured deformation increased with fall angle, from approximately 3.55 mm at 30° to 6.10 mm at 60° and 9.40 mm at 90°. This trend is consistent with the greater impact severity observed at the higher fall angles in the kinematic measurements. Although the maximum deformation increased with fall angle, the compression velocities estimated from the initial loading slope decreased from 24,600 mm/s at 30° to 13,800 mm/s at 60° and 10,800 mm/s at 90°. Thus, the local compression rate and maximum deformation did not vary proportionally across the three fall conditions. We also quantified the volume of the headform undergoing deformation during the impact events using high-speed imaging and DIC (digital image correlation). The estimated deformation volume was 0.000390 m³ for the 30° fall, 0.000460 m³ for the 60° fall, and 0.000645 m³ for the 90° fall. The contact area also increased with fall angle and was approximately 0.00439 m², 0.00713 m², and 0.00791 m² for the 30°, 60°, and 90° falls, respectively. The estimated deformation energy ($U$) was compared with the translational kinetic-energy difference ($\Delta KE$) across the primary impact–rebound event, as shown in Table 3. The deformation energies were $0.69\,J$ for the 30° fall, $2.29\,J$ for the 60° fall, and $6.8\,J$ for the 90° fall.

*Table 3 Comparison of translational kinetic energy and estimated deformation energy for the 30°, 60°, and 90° fall conditions.*

| Energy | 30º | 60º | 90º |
|---|---|---|---|
| Kinetic Energy difference, $\Delta KE$ | 1.63J | 2.85J | 3.59J |
| Strain Energy, $U = \frac{1}{2}F\delta$ | 0.69J | 2.29J | 6.8J |

The closest agreement between the two energy estimates was observed for the 60° fall. Differences were observed for the 30° and 90° conditions, with the estimated deformation energy lower than the kinetic energy at 30° and higher than the kinetic energy at 90°. These differences indicate that simplified energy calculation does not account for all mechanisms of energy transfer and dissipation during impact.

The compression test results were used to characterize the rate-dependent mechanical response of the surrogate headform. Figure 9 shows the force–displacement response obtained at loading rates of 1, 10, 100, and 1000 mm/min, together with the axial and lateral deformation measurements used in the material-property analysis. A characteristic length of 160 mm and a characteristic area of 791 mm² were used in the material-property calculations.

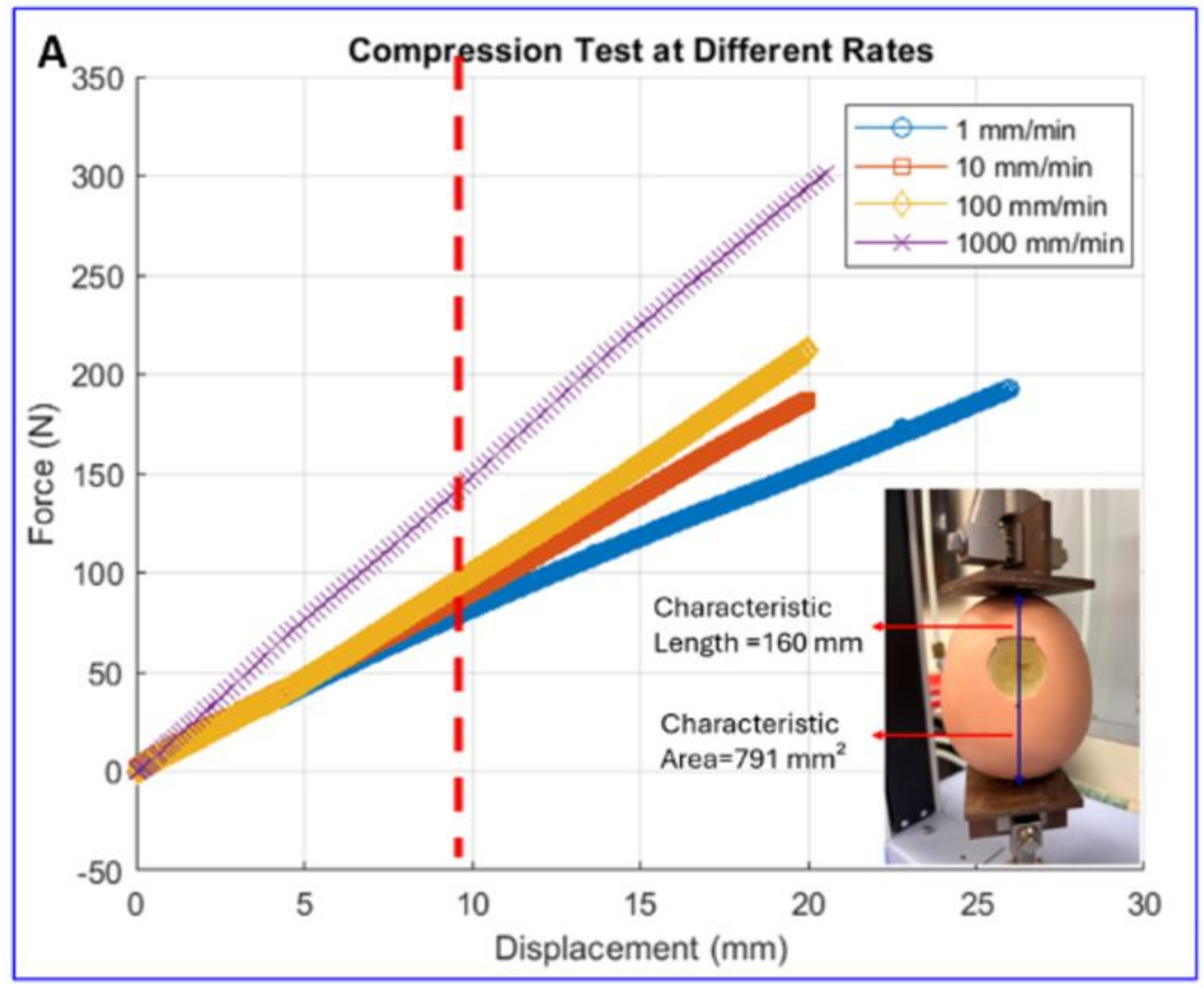

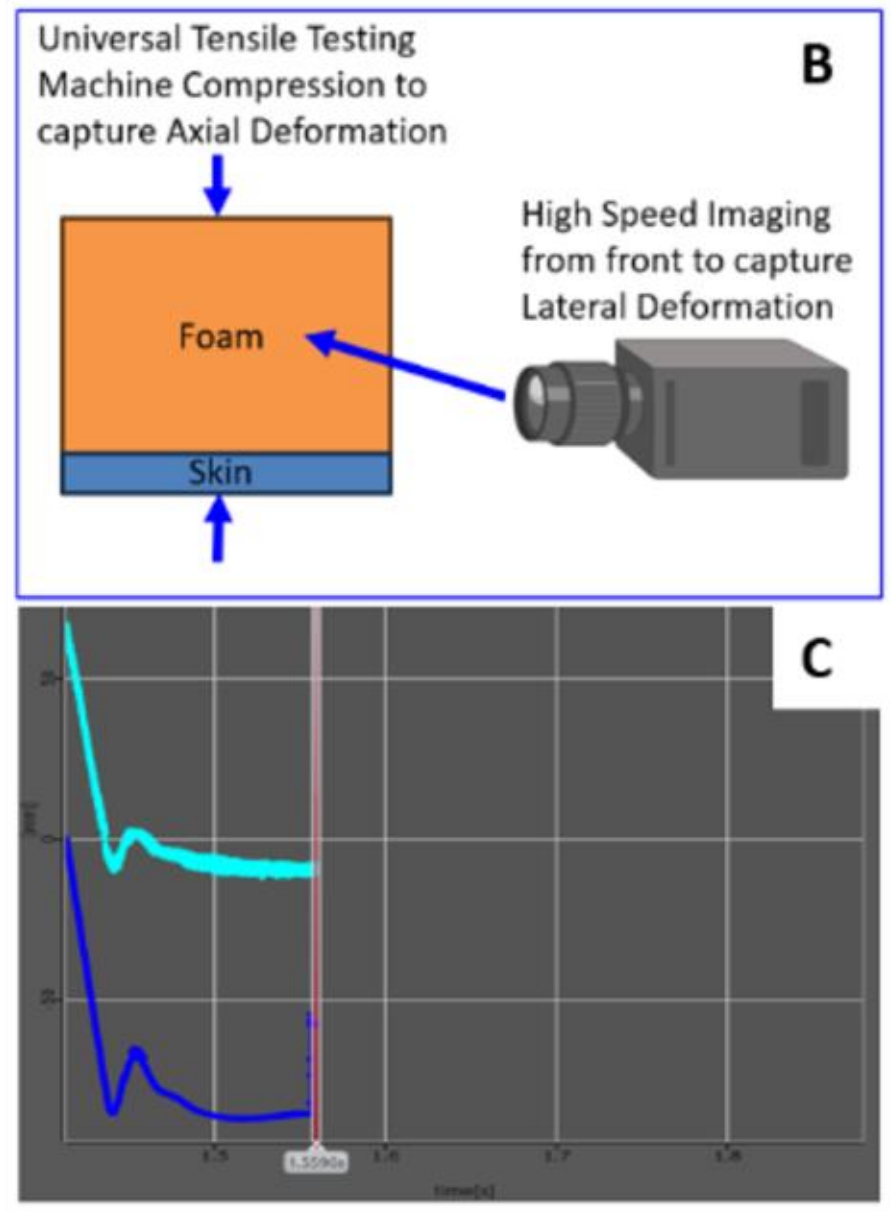

*Figure 9 Characterization of the surrogate headform material response under compression. (a) Force–displacement response measured at loading rates of 1, 10, 100, and 1000 mm/min; the characteristic length and area used in the material-property calculations are also shown. (b) Schematic of the compression test and high-speed imaging arrangement used to measure axial and lateral deformation. (c) Representative displacement histories used to determine the deformation required for calculation of Poisson's ratio.*

Figure 9(a) shows that the force required to produce a given displacement increased with increasing loading rate. The slope of the force–displacement response was greatest at 1000 mm/min and lowest at 1 mm/min, indicating that the surrogate headform exhibits a rate-dependent mechanical response. Using the material-property formulation described in Section 2.1.4, the force-displacement response was used to determine the effective modulus and its dependence on loading rate.

According to conservation of energy, the kinetic energy change and deformation energy would be expected to approach one another if other energy-transfer and dissipative mechanisms were negligible. The closest agreement was observed for the 60° fall, while small differences were observed for the 30° and 90° falls. Therefore, the measured deformation was used together with the kinetic measurements to connect the kinematic response to the contact-mechanics analysis.

Using the deformation measurements and elastic contact model described in Section 2.1.4, the effective modulus was estimated and compared with the compression-test response of the headform. Similar compression tests have been performed previously to characterize the mechanical and rate-dependent behavior of foam materials [30]. The characteristic area used for the material-property calculation was 791 mm² (0.000791 m²), as shown in Figure 9(a). Using this characteristic area and length together with the gradient of the compression-test curve at the highest loading rate, the effective modulus was approximately $3034\ kPa$.

$$E_{SURROGATE\ HEAD} = E_1 = \frac{(\frac{150.23 - 0}{791})}{(10 - 0)/160} = \ 3034\ kPa$$

It is evident from Figure 9a that the surrogate head exhibited a rate-dependent response. Fitting the rate-dependent relationship described in Section 2.1.4 resulted in a reference modulus of $C = 3587\ kPa$ and a rate exponent of $m = 0.0984$.

The effective radius of curvature for the head plate contact was 89 mm. Using this radius together with the experimentally determined effective modulus and Poisson's ratio resulted in a Hertzian contact stiffness of approximately $2x10^6 Nm^{-1.5}$. This experimentally determined stiffness was subsequently used as the elastic component of the Hunt–Crossley contact formulation in the OpenSim model.

### 3.4 BIOMECHANICAL ANALYSIS

The experimentally measured surrogate response was compared with the OpenSim musculoskeletal model to evaluate differences in head–neck motion and head-impact kinematics. The experimentally determined Hertzian contact stiffness described in Section 3.3 was used in the Hunt-Crossley contact formulation of the OpenSim model. The comparison focused on the relative head–neck motion, translational acceleration, and velocity for the three fall conditions. This study directly compares the response of the biofidelic surrogate with the biomechanical response of a human digital twin for the same impact scenario.

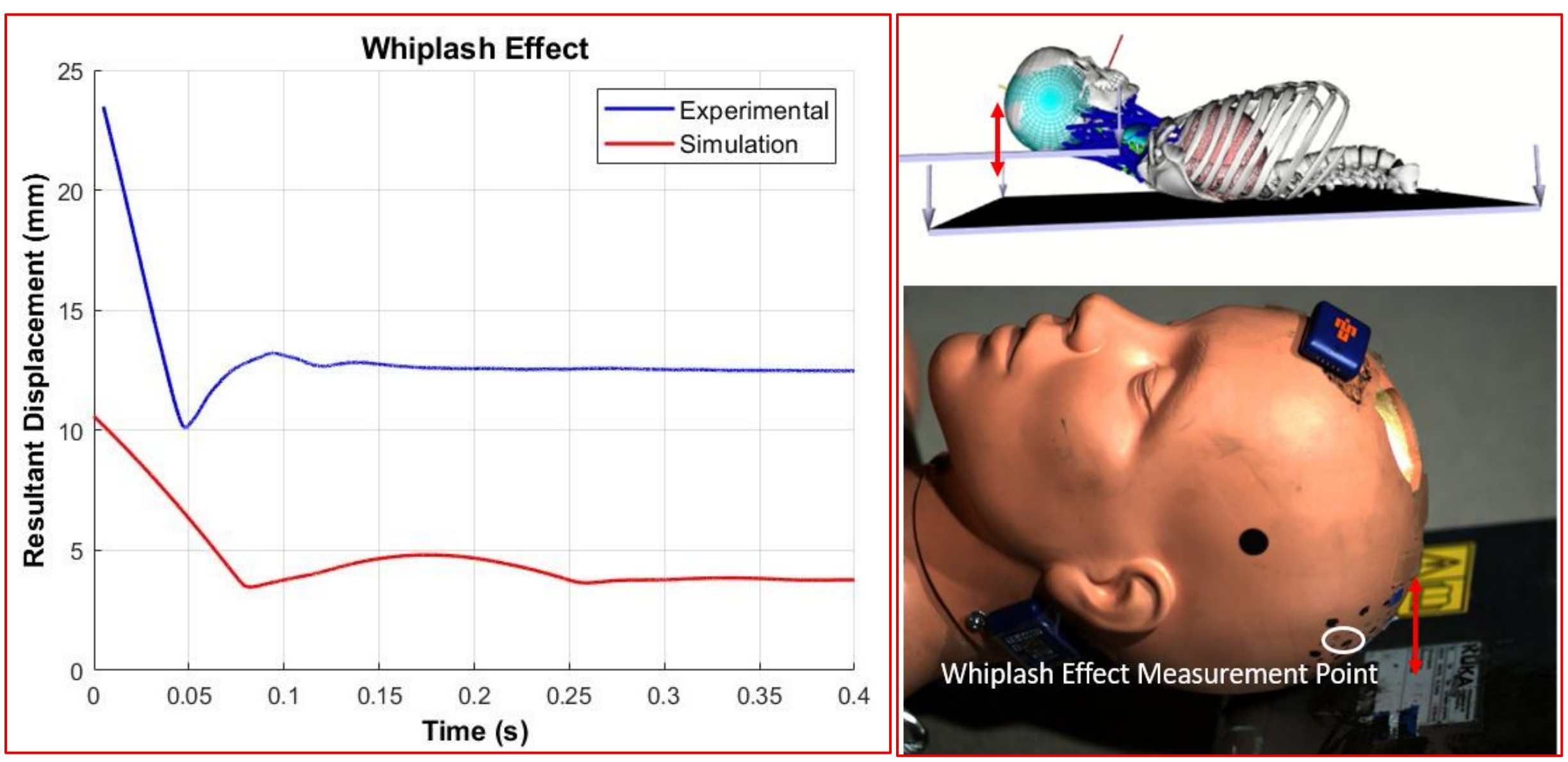

*Figure 10 Whiplash effect time-history comparison between the OpenSim simulation and physical surrogate. (a) Comparison of the experimental and simulated relative head–neck displacement. (b) Method used to quantify the relative head–neck displacement from the OpenSim simulation and high-speed imaging of the physical surrogate.*

Figure 10 compares the relative head–neck motion measured from the surrogate experiment with that predicted by the OpenSim model. The time histories of the two systems were different. The surrogate response (blue line - experimental) showed a large initial peak followed by smaller secondary oscillations, whereas the OpenSim response (red line - simulation) showed lower-amplitude motion that persisted over a longer time. The magnitude of the measured relative head–neck displacement was also greater for the surrogate than for the simulation. This difference may be related to the compliant foam and outer skin of the surrogate headform, whereas the OpenSim head is inherently represented as a rigid body.

In addition to the relative head–neck motion, the translational kinematic response of the two systems was compared for the 30°, 60°, and 90° fall conditions. Figure 11 and Figure 12 illustrate the resultant translational acceleration and velocity time histories for the head across all three fall configurations, comparing the experimental acceleration measurements with the corresponding OpenSim simulation outputs. In Figure 11, for resultant translational acceleration, the overall waveform shape from simulation and experiment is qualitatively similar, with both signals showing sharp spikes associated with impact in each fall. Nonetheless, the accelerometer consistently produces much larger peak resultant accelerations than the simulation model, particularly in the higher-severity falls; the 90° and 60° cases show experimental acceleration peaks several times higher than the corresponding OpenSim peaks, whereas the 30° fall exhibits closer agreement in magnitude.

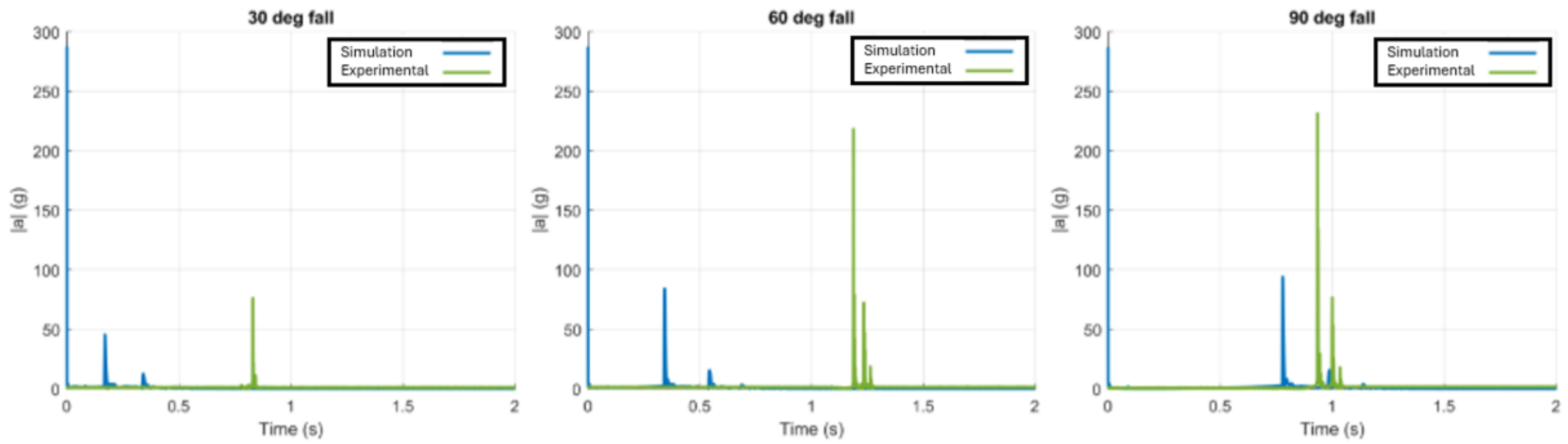

*Figure 11 Comparison of resultant accelerations for all cases*

In Figure 12, Musculoskeletal model simulation typically predicts slightly higher peak resultant velocities than experimental velocities for the 30° and 60° falls, while agreement improves for the 90° fall where both systems report similar peak values. The peak-summary plot reflects this pattern: the regression slope below one, combined with a positive intercept, indicates that the physical surrogate model tends to underestimate smaller peaks but converges toward the OpenSim magnitudes as severity increases. The peak-summary scatter for acceleration confirms a strong overestimation by accelerometers, with a regression slope clearly greater than one and a large negative intercept. When magnitudes and waveform shapes are considered together, the results indicate that the musculoskeletal model provides conservative estimates of peak translational acceleration and slightly higher peak translational velocities relative to the physical surrogate model in lower-severity falls. The systematic biases in peak magnitude imply that relating the musculoskeletal response with the surrogate model response requires parametric calibration and mapping before using for absolute threshold-based injury criteria.

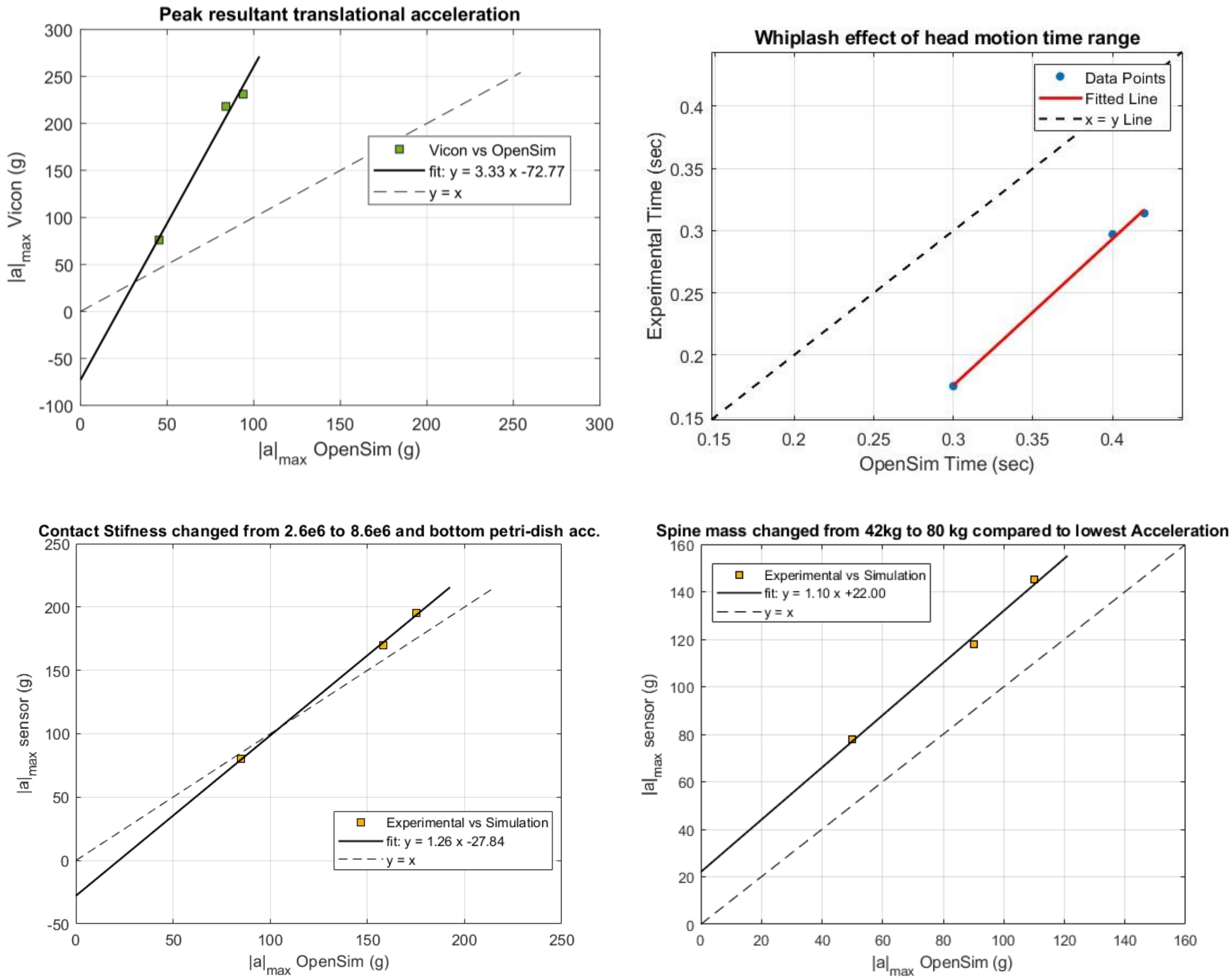

*Figure 12 Comparison of absolute maximum kinematics between experiment and simulation data*

We explore different body mass ratios and contact parameters to understand how they affect head kinematics during fall-induced head collision injury. The purpose of these tests is to identify the key parameters that strongly affect head kinematics as well as head injury. This parameter can then be calculated for a real human head, and surrogate models can be tuned.

From OpenSim simulation results, we observe that increasing head mass leads to reduced acceleration of the center of mass of the head. Also, an increase in spine mass leads to increased acceleration but saturates after 80 kg. Further increase in mass leads to no further acceleration increase of the head form. Hence, the effect of body mass ratio is negligible.

The sharpest acceleration change comes from increasing the Hertz Contact stiffness between the impactor plate and the head model. Analyzing the Hertz contact force interaction of human subjects and designing surrogate models accordingly will nullify the need for further human subject involvement.

Hence, the parameter of interest to tune is Hertz Contact stiffness. We compare the head acceleration between simulation and experiment at different contact stiffnesses as shown in Figure 12 to obtain a correlation factor.

For experimental and OpenSim acceleration:

$$|a|_{max}^{experimental} = \alpha_a . |a|_{max}^{opensim} + \beta_a \tag{9}$$

From our data, $\alpha_a = 3.33$, and $\beta_a = -72.77$.

For experimental vs OpenSim whiplash time:

$$|w|_{max}^{experimental} = \alpha_w . |w|_{max}^{opensim} + \beta_w \tag{10}$$

From our data, $\alpha_w = 1.178$, and $\beta_w = -0.178$

While these coefficients are preliminary and derived from a limited number of impacts, they provide a pragmatic calibration framework for adjusting sensor-based peak kinematics toward the OpenSim reference before using them in quantitative injury risk analyses. The whiplash effect time range for the simulation is much greater than the time range of the experimental data. However, the amplitude of the whiplash effect in the experiment is much greater due to head form deformation, as the head is made of foam inside and thin rubber-like material outside. The simulation results consider human head dynamics, including skull deformation, leading to very small whiplash deformations of the head during multiple impacts of the head with the steel plate. Also, the head-neck interaction plays a key role in the whiplash effect.

It is worth noting that in an OpenSim model, the head is represented as a rigid body with a prescribed mass, center of mass, and mass moment of inertia. Its motion is determined by the overall musculoskeletal dynamics of the body, including the motion of individual body segments, joint constraints, muscle forces, and the forces and moments transmitted through the neck. These interactions determine the resulting linear and angular acceleration of the head during impact. When the rigid head approaches and contacts a rigid plate, in principle, the generated contact force should be zero as the contact deformation would be zero. In reality, when contact between two rigid bodies is simulated by an OpenSim model, neither body is physically deformed. Instead, the contact model uses the relative geometric penetration between the two bodies as a measure of contact deformation. As the head moves toward the rigid plate, the distance between the contact surfaces is continuously calculated. Once the surfaces mathematically overlap, the amount of overlap is treated as the relative deformation or penetration. This numerical penetration is then used by the contact model to calculate the contact force (Eqn. (1)). Therefore, $\delta$ represents a virtual deformation used to model contact compliance, rather than actual deformation of either rigid body. As such, the presented correlation factor should be treated as a modified contact displacement to allow the rigid head of an OpenSim model to exhibit realistic contact deformation.

## 3.5 LIMITATIONS

The present study has several limitations that should be considered when interpreting the results. First, the surrogate headform does not reproduce the anatomical structure or material properties of the human head. The surrogate consists primarily of a compliant foam core and outer skin, whereas the human head contains a rigid skull surrounding heterogeneous and viscoelastic brain tissue. This difference is particularly important for the deformation and contact response and likely contributes to some of the differences observed between the experimental surrogate and the OpenSim model. The OpenSim model also represents the head as a rigid body and therefore does not capture the local deformation and regional response measured in the physical surrogate. Future studies could use a more biofidelic headform with separate skull, brain, and soft-tissue components and compare it with a deformable computational head model.

The experimental fall conditions were also limited to seated backward falls at 30°, 60°, and 90° onto a rigid steel plate, with three repetitions performed for each condition. Therefore, the results should not be generalized to all fall configurations or impact surfaces. Future studies should include additional fall angles, impact locations, surface stiffnesses, and other relevant factors. Some of the measured accelerations also approached the ±200g range in each axis of the IMUs, and higher-range sensors could be used in future high-severity tests to reduce the possibility of measurement saturation.

Another limitation is associated with the material characterization and contact model. The compression tests were performed at loading rates up to 1000 mm/min, whereas the deformation velocities measured during impact were substantially higher. Therefore, application of the fitted rate-dependent relationship to the actual impact condition requires extrapolation beyond the experimentally tested loading-rate range. Future work should include high-rate material testing that more closely represents the deformation rates measured during the fall impacts. In addition, Hertz contact theory was used as an initial elastic description of the head-plate interaction even though the surrogate headform is layered, deformable, and viscoelastic. Direct measurement of head-plate contact force using a force plate or force-torque sensor would provide an additional means of validating the estimated contact stiffness and the Hunt-Crossley contact parameters.

The energy comparison also revealed differences between the translational kinetic-energy change and the estimated deformation energy, particularly for the 30° and 90° falls. This indicates that the simplified analysis does not capture all pathways of energy transfer, which may include rebound, rotational motion, viscoelastic dissipation, head-neck motion, and deformation of other parts of the surrogate. Future studies could combine force measurements, three-dimensional deformation measurements, and full-body kinematics to develop a more complete impact energy balance.

Finally, the SH-SY5Y cells were cultured in two-dimensional Petri dishes placed within the surrogate headform. Although this approach allows cellular response to be studied together with the local mechanical loading, it does not reproduce the three-dimensional cellular organization, extracellular matrix, or heterogeneous structure of brain tissue. The measured ROS, viability, and morphological responses should therefore be interpreted as early cellular indicators rather than direct measures of human brain injury. Future studies could incorporate three-dimensional neural cultures, organoids, or other tissue-like models and evaluate additional cellular responses over longer post-impact time periods.

# 4 CONCLUSION:

The study of brain injury due to a mechanical impact is a complex multidisciplinary event where several aspects need to be considered to fully comprehend the outcomes. This includes fall motion kinematics, head deformation at the point of impact, force transmitted to the internal head model, and the response of brain cells situated inside the model due to the impact, and how these results from a surrogate model compare to a real human model response. In this study, we aim to cover all aspects of the impact event and correlate and validate responses from different perspectives. We observed that the regional acceleration response increased with increasing fall angle, together with an increase in headform deformation and changes in cellular response. The median cell-pack peak acceleration was approximately 60g for the 30° fall, 211g for the 60° fall, and 236g for the 90° fall. Under the same conditions, the maximum headform deformation increased from approximately 3.55 mm at 30° to 6.10 mm at 60° and 9.40 mm at 90°. At the 30° fall, the cellular displacement was relatively diffuse, and the changes in oxidative stress and viability were comparatively small, while the 90° condition showed cell drift was clumped in the same location, inferring focal injury, together with the most prominent increase in oxidative stress and reduction in cell viability. These results indicate that increasing impact severity is associated not only with greater acceleration but also with greater headform deformation and changes in cellular response. Regions away from the line of impact force experience less accelerative force, and less damage is expected to occur to neuron cells. The regional differences observed within the same fall condition further show that a single acceleration measurement from one location may not adequately characterize the mechanical loading experienced throughout the headform. Therefore, multi-regional acceleration measurements, together with deformation and cellular response, provide a more complete characterization of the surrogate response during head impact. We also observe multiple waves of dynamic acceleration experienced by the surrogate head at each point. The multiple waves can be explained by the whiplash effect captured by the high-speed imaging, which also shows that this complex motion profile is due to head and neck interaction rather than a simple pendulum fall. Hence, we choose the whiplash effect as a parameter of comparison between the humanoid model and OpenSim model. We see that the correction factor between the two models is an offset of 1.178 and a gradient difference of only 0.178. The OpenSim model considers musculoskeletal aspects, and hence the surrogate model can be used to measure the whiplash effect, and adding this correction factor can bring the results more towards a humanoid head-neck response. Also, we choose linear acceleration due to regional differences as a point of comparison. From the correction factors between the surrogate and musculoskeletal head models, we see that the acceleration factor has more offset between the surrogate and simulation results. This difference is because OpenSim considers the head form as a rigid body which is not a true comparison for the soft brain tissue component which has regional differences. We feel that a central factor in developing a surrogate model toward a more humane model is replacing the head form with a more realistic soft tissue that mimics the brain organ, even though, compared to an OpenSim model, our surrogate system predicts brain anatomy and regional differences better. This capability of the surrogate system to capture regional differences is highly important in the study of brain injury biomechanics.

The novelty of this study is an integrated platform for linking macro-level impact kinematics and contact mechanics to cellular-level response. Although the current framework includes simplifications in both the physical surrogate and musculoskeletal model, it provides a controlled and repeatable approach for examining how fall conditions, regional acceleration, head deformation, and contact properties are associated with cellular response. The framework can therefore support future development of more biofidelic surrogate models by identifying mechanical parameters, particularly contact stiffness, that can be adjusted and evaluated against human-relevant head impact behavior.

**Authors Contribution Statement:**

R.A. performed the material testing using the universal testing machine (UTM), conducted the high-speed camera motion tracking experiments, performed the cell culture experiments, analyzed the biological results, and carried out the deformation analysis. M.I.H. performed the surrogate fall experiments and relevant motion analysis. R.Z. performed the numerical simulations, analyzed the simulation results and wrote this segment of the work. All authors contributed to the interpretation of the results and writing the initial draft of the manuscript. A.A. designed the study, acquired funding, and revised the manuscript. All the authors reviewed the manuscript and approved the current version.

## Acknowledgement:

A.A. acknowledges the financial support from Office of Naval Research (ONR)'s DURIP program (N00014-23-1-2232) and Force Health Protection (FHP) program (through the Awards # ONR: N000142512331 and ONR: N000142412324: Dr. Timothy Bentley, Program Manager).